\documentclass[twoside,11pt]{article}
\renewcommand{\baselinestretch}{1.3}

\usepackage{amssymb,amsthm} 

\newtheorem{defn}{Definition}
\newtheorem{lemma}{Lemma}
\newtheorem{prop}{Proposition}
\newtheorem{cor}{Corollary}
\newtheorem{thm}{Theorem}

\theoremstyle{definition}
 
\newtheorem{ex}{Example}
\newtheorem{remark}{Remark}

\title{\bf Correspondence theorems for hierarchies of equations of pseudo-spherical type}
 
\author{Enrique G. Reyes}
 
\date{Department of Mathematics \\
      University of Oklahoma \\  
      Norman, OK 73019, USA \\
      E-mail: ereyes@math.ou.edu}
 
\begin{document}
\maketitle
\begin{abstract}
Hierarchies of evolution equations of pseudo-spherical type are introduced, generalizing 
the notion of a single equation describing pseudo-spherical surfaces due to S.S. Chern and K.
Tenenblat, and providing a connection between differential geometry and the study of hierarchies 
of equations which are the integrability condition of $sl(2,{\bf R})$--valued linear problems. 
As an application, it is shown that there exists a local correspondence between {\em any two} 
(suitably generic) solutions of arbitrary hierarchies of equations of pseudo-spherical type.
\end{abstract}

\pagestyle{myheadings}

\markboth{REYES\hspace{70mm}}{\hspace{20mm}CORRESPONDENCE THEOREMS FOR HIERARCHIES}


\section{Introduction}     

The class of equations of pseudo-spherical type (or ``describing pseudo-spherical surfaces'') 
was introduced by S.S. Chern and K. Tenenblat \cite{6} in 1986, motivated by the following 
observation by Sasaki \cite{20}:
the graphs of generic ---in a sense to be made precise in Section 2--- solutions of equations 
integrable by the Ablowitz, Kaup, Newell, and Segur (AKNS) inverse scattering approach can be
equipped, whenever their associated linear problems are real, with Riemannian metrics of 
constant Gaussian curvature equal to $-1$. Chern and Tenenblat then called an equation $\Xi = 0$ 
of {\em pseudo-spherical type} if the graphs of generic solutions of $\Xi = 0$ can be 
equipped with Riemannian (or Lorentzian) metrics of constant Gaussian curvature $-1$. The 
precise definition is given in Section 2.

The importance of this structure for differential equations, and specially for the theory of 
integrable systems, arises from several facts. First, some of their fundamental properties, 
such as the existence of conservation laws, symmetries, and B\"{a}cklund transformations, can 
be understood by geometrical means \cite{brt,6,17,175,19,20,T}. Second, these equations are the 
integrability condition of $sl(2,{\bf R})$--valued linear problems (Sasaki \cite{20}, Chern and 
Tenenblat \cite{6} and Section 2 below), and therefore one can try to obtain solutions for them 
using a scattering/inverse scattering approach \cite{brt}. Third, one can obtain new analytical 
results motivated by geometrical characteristics of pseudo-spherical surfaces. For example, 
inspired by the well--known observation that two surfaces of constant Gaussian curvature equal 
to $-1$ are locally indistinguishable, Kamran and Tenenblat \cite{12} and the present author 
\cite{re} have showed that, if $\Xi = 0$ and $\widehat{\Xi} = 0$ are of pseudo-spherical type 
and $u(x,t)$ and $\widehat{u}(\widehat{x}, \widehat{t}\,)$ are generic solutions of $\Xi = 0$ 
and $\widehat{\Xi} = 0$ respectively, one can relate $u(x,t)$ and $\widehat{u}(\widehat{x}, 
\widehat{t}\,)$ by integrating {\em first order} systems of equations ---thereby strongly 
extending B\"{a}cklund's \cite{BA,T} theorem--- {\em and} one can obtain a {\em formula} for 
$\widehat{u}(\widehat{x},\widehat{t}\,)$ in terms of the function $u(x,t)$. 

Now, an aspect of the theory of integrable systems which appears to have been little studied 
from a geometrical point of view is the fact that, as stressed for instance, by Faddeev and 
Takhtajan \cite{8}, evolution equations $u_{t} = F$ which are the integrability condition of 
a non--trivial one--parameter family of linear problems 
\begin{equation}
v_{x} = X\, v \; , \; \; \; \; \; \; \; \; \; \; v_{t} = T\, v \; ,   \label{l1}
\end{equation}
in which $X$ and $T$ are, say, $sl(2,{\bf R})$--valued functions of $u$ and a finite number of
its derivatives, are members of {\em infinite hierarchies} of evolution equations $u_{\tau_{n}} 
= F_{n}$ possessing the following characteristics: {\em (a)} they generate a hierarchy of 
commuting flows, {\em (b)} they are also integrability conditions of a non--trivial 
one--parameter families of linear problems, and {\em (c)} they share with the given equation the
``space'' part of the linear problem (\ref{l1}), that is, their associated linear problems are 
of the form
\[
v_{x} = X\, v \; , \; \; \; \; \; \; \; \; \; \; v_{\tau_{n}} = T_{n}\, v \; .
\] 

One wonders if these facts have counterparts in the class of equations considered by 
Chern and Tenenblat: can one can define hierarchies of evolutionary 
equations describing pseudo-spherical surfaces, and do these ``hierarchies of pseudo-spherical 
type'' possess the characteristics {\em (a), (b)} and {\em(c)} mentioned in the last paragraph? 
Moreover, can one generalize to this new setting the correspondence theorems between solutions 
of single equations of pseudo-spherical type \cite{12,re} which were cited above?
This paper is devoted to answering these two questions. Hierarchies of pseudo-spherical type
are introduced in Section 3 (a preliminary definition appeared in \cite{195}) and it is shown 
that they do satisfy {\em (a), (b)} and {\em (c)} above. In particular, they do generate
families of mutually commuting flows. Next, it is proven in Section 4 that there exist 
correspondences between solutions of {\em any two} hierarchies of equations of pseudo-spherical 
type, generalizing the results of \cite{12,re}. As an example, it is shown at the end of Section
4 how one can relate solutions of the KdV hierarchy to solutions of a hierarchy of linear 
equations. 

Some of the results appearing in \cite{re} and this paper were announced at the 2000 NSF--CBMS
Conference on the Geometrical Study of Differential Equations \cite{howard} and more recently 
in \cite{letter}.

\section{Equations of pseudo-spherical type}

\begin{defn}
A two-dimensional manifold $M$ is called a pseudo-spherical surface if there exist one-forms
$\overline{\omega}^{1}, \overline{\omega}^{2}, \overline{\omega}^{3}$ on $M$ that satisfy the
independence condition $\overline{\omega}^{1}\wedge \overline{\omega}^{2}\ne 0$,
and the structure equations
\begin{equation}               \label{structure0}
d\,\overline{\omega}^{1} = \overline{\omega}^{3} \wedge \overline{\omega}^{2} \; ,  
\; \; \; \; \; \; \; 
d\, \overline{\omega}^{2} = \overline{\omega}^{1} \wedge \overline{\omega}^{3} \; , 
\; \; \; \; \; \; \; 
d\, \overline{\omega}^{3} = \overline{\omega}^{1} \wedge \overline{\omega}^{2} \; .
\end{equation}
\end{defn}

Thus, if $M$ is pseudo-spherical, it is equipped with the Riemannian metric $ds^{2} =
(\overline{\omega}^{1})^{2} + (\overline{\omega}^{2})^{2}$, the corresponding torsion--free
metric connection one--form is $\overline{\omega}^{3}$, and its Gaussian curvature is $-1$.

Hereafter, partial derivatives $\partial^{n+m}u/\partial x^{n} \partial t^{m}$, $n,m \geq 0$, 
are denoted by $u_{x^{n}t^{m}}$.

\begin{defn}         \label{epss}
A scalar differential equation    \begin{equation}
\Xi (x, t, u, u_{x}, \dots, u_{x^{n}t^{m}}) = 0 \; ,  \label{evu}
\end{equation}
in two independent variables $x,t$ is of pseudo-spherical type (or, it describes pseudo-spherical
surfaces) if there exist one-forms $\omega^{\alpha}$,
\begin{equation}   \label{omega}
\omega^{\alpha} = f_{\alpha 1}(x, t, u, \dots, u_{x^{r}t^{p}})\,dx +
f_{\alpha 2}(x, t, u, \dots, u_{x^{s}t^{q}})\,dt \; ,                    \; \; \; \;
\; \alpha = 1,2,3
\end{equation}
whose coefficients $f_{\alpha\beta}$ are differential functions, such that the one--forms
$\overline{\omega}^{\alpha} = \omega^{\alpha}(u(x,t))$  satisfy the structure equations 
$(\ref{structure0})$ whenever $u=u(x,t)$ is a solution to Equation $(\ref{evu})$.
\end{defn}

Recall that a differential function is a smooth function which depends on $x,t$, and a finite 
number of derivatives of $u$ \cite{16}; the trivial case when the functions 
$f_{\alpha\beta}$ all depend only on $x,t$, is excluded from the considerations below. 

\begin{ex}
Burgers' equation $u_{t} = u_{xx} + u u_{x} + h'(x)$ is an equation of 
pseudo-spherical type. Associated one--forms $\omega^{\alpha}$ are
\begin{eqnarray*}
\omega^{1} & = & \left( \frac{1}{2} u - \frac{\beta}{\eta} \right) dx +
                 \left( \frac{1}{2} u_{x} + \frac{1}{4} u^{2} + \frac{1}{2} h(x) \right) dt \; , 
                                                                                  \label{o1} \\
\omega^{2} & = & - \omega^{3} \; = \; \eta\,dx + \left( \frac{\eta}{2} u + \beta \right) dt \; ,
\end{eqnarray*}
in which $\eta \neq 0$ is a parameter, and $\beta$ is a solution of the equation
$\beta^{2} - \eta \beta_{x} + (\eta^{2}/2) h(x) = 0$.
\end{ex}

The expression ``PSS equation'' will be sometimes utilized below instead of the more formal 
phrase ``equation of pseudo-spherical type''. 

\begin{defn}
Let $\Xi = 0$ be a PSS equation with associated one--forms
$\omega^{\alpha}$, $\alpha=1,2,3$. A solution $u(x,t)$ of $\Xi = 0$ is $I$--generic if
$(\omega^{3}\wedge\omega^{2})(u(x,t)) \neq 0$, $II$--generic if
$(\omega^{1}\wedge\omega^{3})(u(x,t)) \neq 0$, and $III$--generic if
$(\omega^{1}\wedge\omega^{2})(u(x,t)) \neq 0$.
\end{defn}
 
For instance (see Example 1) the traveling wave $u(x,t) = 2 {\rm e}^{x+t}/(1+{\rm e}^{x+t})$ is 
a $III$--generic solution of Burgers equation if $h(x) = 0$, but it is not $I$--generic. These 
genericity notions allow one to interpret Definition
2 geometrically as follows:
 
\begin{prop}
Let $\Xi = 0$ be a PSS equation with associated one--forms $\omega^{\alpha}$, let $u(x,t)$ be a
solution of $\Xi = 0$, and set $\overline{\omega}^{\alpha} = \omega^{\alpha}(u(x,t))$.

{\em (a)} If $u(x,t)$ is $I$--generic, $\overline{\omega}^{2}$ and
$\overline{\omega}^{3}$ determine a Lorentzian metric of Gaussian curvature $K = -1$ on the
domain of $u(x,t)$, with connection one--form given by $\overline{\omega}^{1}$.

{\em (b)} If $u(x,t)$ is $II$--generic, $\overline{\omega}^{1}$ and
$- \overline{\omega}^{3}$ determine a Lorentzian metric of Gaussian curvature $K = -1$ on
the domain of $u(x,t)$, with connection one--form given by $\overline{\omega}^{2}$.

{\em (c)} If $u(x,t)$ is $III$--generic, $\overline{\omega}^{1}$ and
$\overline{\omega}^{2}$ determine a Riemannian metric of Gaussian curvature $K =
-1$ on the domain of $u(x,t)$, with connection one--form given by $\overline{\omega}^{3}$.
\end{prop}   

Proposition 1 is proven in \cite{re}. It follows from considering the structure equations of a 
surface equipped with a  metric of signature $(1,\epsilon )$, $\epsilon = \pm 1$ which appear,
for example, in \cite{T}.

The invariance properties of the structure equations (\ref{structure0}) are spelled out in the 
following straightforward proposition:
\begin{prop}   \label{p1}
Let $\omega^{\alpha}$, $\alpha = 1,2,3$, be one--forms whose coefficients are differential 
functions. Let $u(x,t)$ be a smooth function, and set $\overline{\omega}^{\alpha} = 
\omega^{\alpha}(u(x,t))$. The structure equations
\begin{equation}
d\,\overline{\omega}^{1} = \overline{\omega}^{3} \wedge \overline{\omega}^{2} \; , 
\; \; \; \; \; \; \; \; 
d\,\overline{\omega}^{2} = \overline{\omega}^{1} \wedge \overline{\omega}^{3} \; , 
\; \; \; \; \; \; \; \;
\mbox{ and } \; \; \; \; \; \; \; \;
d\,\overline{\omega}^{3} = \overline{\omega}^{1} \wedge \overline{\omega}^{2} \;  , 
                                                                             \label{structure1}
\end{equation}
are invariant under the transformations
\begin{equation}
\widehat{\omega}^{1}  = \overline{\omega}^{1} \cos \overline{\rho} + \overline{\omega}^{2} 
\sin \overline{\rho} \; , ~ ~ ~ ~ ~ ~ ~
  \widehat{\omega}^{2}  =  - \overline{\omega}^{1} \sin \overline{\rho} + \overline{\omega}^{2} 
\cos \overline{\rho} \; , ~ ~ ~ ~ ~ ~ ~
  \widehat{\omega}^{3}  =  \overline{\omega}^{3} + d \overline{\rho} \; ;  \label{I}
\end{equation}
\begin{equation}
 ~ ~ \widehat{\omega}^{1}  =  \overline{\omega}^{1}\cosh \overline{\rho} - \overline{\omega}^{3}
\sinh{\rho} \; ,  ~ ~ ~
\widehat{\omega}^{2}  =  \overline{\omega}^{2} + d \overline{\rho} \; , ~ ~ ~ ~ ~ 
\widehat{\omega}^{3}  =  -\overline{\omega}^{1} \sinh \overline{\rho} + \overline{\omega}^{3} 
\cosh \overline{\rho} \; ; \label{II}
\end{equation}
and
\begin{equation}     
~ ~ \widehat{\omega}^{1}  = \overline{\omega}^{1} + d \overline{\rho} \; , ~ ~ ~ ~
  \widehat{\omega}^{2}  = \overline{\omega}^{2} \cosh \overline{\rho} + \overline{\omega}^{3} 
\sinh \overline{\rho} \; , ~ ~ ~ ~ ~ ~ 
  \widehat{\omega}^{3}  =  \overline{\omega}^{2} \sinh \overline{\rho} + \overline{\omega}^{3} 
\cosh \overline{\rho} \; ;  \label{III}
\end{equation}
in which $\rho$ is any differential function and $\overline{\rho} = \rho(u(x,t))$.
\end{prop}

The geometric interpretation of Equations (\ref{I})--(\ref{III}) follow from Proposition 1.
Indeed, if $\Xi = 0$ is a PSS equation with associated one--forms 
$\omega^{\alpha}$, $\alpha=1,2,3$, and $u(x,t)$ is a $III$--generic solution of $\Xi = 0$, 
(\ref{I}) is simply the transformation induced on the one--forms $\overline{\omega}^{\alpha}$, 
$\alpha=1,2,3$ by a rotation of the moving orthonormal frame dual to the coframe 
$\{ \overline{\omega}^{1}, \overline{\omega}^{2} \}$. Analogously, if $u(x,t)$ is $II$--generic,
(\ref{II}) is the transformation induced on the one--forms 
$\overline{\omega}^{\alpha}$ by a Lorentz boost of the moving frame dual to the coframe 
$\{ \overline{\omega}^{1}, -\overline{\omega}^{3} \}$, and if $u(x,t)$ is $I$--generic, 
(\ref{III}) is the transformation induced on the one--forms $\overline{\omega}^{\alpha}$ by a 
Lorentz boost of the moving frame dual to the coframe $\{ \overline{\omega}^{2}, 
\overline{\omega}^{3} \}$.   

\begin{remark}
A very careful analysis of the invariance properties of pseudo--Riemannian surfaces of constant 
scalar curvature and their relation with classical B\"{a}cklund transformations has been made by 
Crampin, Hodgkin, Robinson, and McCarthy in \cite{CHRM}.   
\end{remark}

As pointed out in the Introduction, equations of pseudo-spherical type are of interest because, 
for instance, their conservation laws, symmetries, and classical B\"{a}cklund transformations 
can be studied from a geometrical point of view \cite{brt, 6, 17, 175, 19, 20, T}. For this 
paper it is important to stress that if $\Xi = 0$ describes pseudo-spherical surfaces, it is 
the integrability condition of a $sl(2,{\bf R})$--valued linear problem. In fact, it is not hard
to see that if $\Xi(x,t,u,\dots) = 0$ is a PSS equation with associated one--forms 
$\omega^{\alpha}$, $\alpha=1,2,3$, the $sl(2,{\bf R})$--valued linear problem
\begin{equation}
\frac{\partial v}{\partial x} = X v \; , 
\, \, \, \, \, \, \, \, \, \, \, \, \, \, \, \, \, \, \, \,\, \, \,
\frac{\partial v}{\partial t} = T v \; ,                     \label{linear1}
\end{equation}
in which $X$ and $T$ are determined by the $sl(2,{\bf R})$--valued one--form
\begin{equation}
\Omega = X dx + T dt     = \frac{1}{2} \left(
                          \begin{array}{lr}
                                 \omega^{2}& \omega^{1} - \omega^{3}\\
                                 \omega^{1} + \omega^{3}&  - \omega^{2}
                          \end{array} \right) \; ,                              \label{linear2}
\end{equation}
is integrable whenever $u(x,t)$ is a solution to $\Xi = 0$. In other words, the structure 
equations (\ref{structure0}) imply that the matrix equation
\begin{equation}
\frac{\partial X}{\partial t} - \frac{\partial T} {\partial x} +  [X,T] = 0        \label{zcc}
\end{equation}
is identically satisfied whenever $u(x,t)$ is a solution of $\Xi=0$. It follows then that one 
may hope to study PSS equations via scattering/inverse scattering techniques. An interesting 
example is provided by the equation
\[
\{ u\sb {t} - [ \alpha g(u) + \beta ] u\sb {x} \}\sb {x} =  g'(u) \; ,   \label{rabelo}
\]
in which $g(u)$ satisfies $g''+\mu g=\theta$, and $\mu, \theta , \alpha , \beta$ are real 
numbers. M. Rabelo proved in 1989 (see \cite{brt,T}) that (\ref{rabelo}) is of pseudo-spherical 
type with associated one--forms  
\[
\omega^{1} = \zeta u_{x}\,dx + \zeta (\alpha g + \beta ) u_{x}\,dt \, , \; \; \; \; \;
\omega^{2} = \eta\,dx + ( (\zeta^{2} g - \theta)/\eta + \beta\eta )\,dt \, \; \; \; \; \;
\omega^{3} = (\zeta g'/\eta)\,dt \; ,
\]
in which $\zeta^{2} = \alpha \eta^{2} - \mu$. Beals, Rabelo and Tenenblat \cite{brt} then used
the corresponding linear problem (\ref{linear1}) and (\ref{linear2}) to solve (\ref{rabelo}) by 
scattering/inverse scattering methods.

\begin{remark}
It is a classical observation in integrable systems \cite{8} that one can interpret Equations  
(\ref{linear1})--(\ref{zcc}) in terms of an $SL(2, {\bf R})$ connection: consider a principal 
fiber bundle $U \times SL(2, {\bf R}) \rightarrow U$, in which $U$ is the space of independent 
variables $(x,t)$. Then, the $sl(2, {\bf R})$--valued one--form $\Omega$ defined in 
(\ref{linear2}) ---and satisfying (\ref{zcc}) whenever $u(x,t)$ is a solution to the equation 
$\Xi = 0$--- can be thought of as a flat connection on the principal bundle 
$U \times SL(2,{\bf R}) \rightarrow U$, and the linear problem (\ref{linear1}) as determining 
a covariantly constant section $v(x,t)$ on a trivial vector bundle $U \times V \rightarrow U$ 
associated to the bundle $U \times SL(2,{\bf R})$, in which $V$ is some two--dimensional 
vector space. This interpretation is of importance in the study of transformations of solutions
to equations of pseudo-spherical type \cite{re}.
\end{remark}

Now, an important idea in the study of integrable systems is (Faddeev and Takhtajan \cite{8}) 
that an equation $\Xi = 0$ is not just the integrability condition of a linear problem $dv = 
\Omega v$, but that the ``zero curvature equation'' $X_{t} - T_{x} + [X,T] = 0$ is {\em 
equivalent} to $\Xi =0$. It is precisely under this assumption that nonlinear equations which 
are the integrability condition of linear problems have been found from geometrical 
considerations, see for example the papers by Bobenko \cite{Bo}, Lund and Regge \cite{LR}, 
Lund \cite{L1,L2}, Reyes \cite{175,19}, and the monograph \cite{T} by K. Tenenblat. A crucial 
problem ---which in its full generality is still open--- is to formalize this equivalence within
the context of PSS equations. If $\Xi = 0$ is a $k^{\rm th}$ order evolution equation, 
$\Xi = u_{t} - F(x,t,u, ... , u_{x^{k}})$, one proceeds thus \cite{12,123}:
 
Consider the differential ideal $I_{F}$ generated by the two--forms
\[
du \wedge dx + F(x,t,u, ... , u_{x^{k}}) dx \wedge dt, \; \; \;
du_{x^{l}} \wedge dt - u_{x^{l+1}}\, dx \wedge dt, \; \; \;  1 \leq l \leq
k-1,
\]
on a manifold $J$ with coordinates $x,t,u,u_{x},\ldots ,u_{x^{k}}$. Then, local solutions of
$u_{t} = F$ correspond to integral submanifolds of the exterior differential system $\{ I_{F} ,
dx \wedge dt \}$. One says that $u_{t} = F$ is {\em strictly pseudo-spherical} if the ideal 
$I_{F}$ is algebraically equivalent to a system of differential forms satisfying the 
pseudo-spherical structure equations if pulled back by local solutions of $u_{t} = F$:
 
\begin{defn} \label{strict}
An evolution equation $ u_{t} = F(x,t,u, ... , u_{x^{k}})$ is strictly pseudo-spherical if there
exist one--forms $\omega^{\alpha} = f_{\alpha 1}\,dx + f_{\alpha 2}\,dt$, $\alpha=1,2,3$, whose
coefficients $f_{\alpha\beta}$ are smooth functions on $J$, such that the two--forms
\begin{equation}
\Omega_{1} = d \omega^{1} - \omega^{3} \wedge \omega^{2} \; , \; \; \; \; \; 
\Omega_{2} = d \omega^{2} - \omega^{1} \wedge \omega^{3} \; , \; \; \; \; \; 
\Omega_{3} = d \omega^{3} - \omega^{1} \wedge \omega^{2} \; ,                      \label{mari7}
\end{equation}
generate $I_{F}$.
\end{defn}
 
It follows that if $u_{t} = F$ is strictly pseudo-spherical, it is necessary and sufficient for 
the structure equations $\Omega_{\alpha} = 0$ to hold. Definition \ref{strict} will provide some
important motivation for the constructions appearing in the following section.
The following lemma, proven in \cite{re}, will also be of interest:
 
\begin{lemma}   \label{lemma0}
Necessary and sufficient conditions for the $k^{\rm th}$ order evolution equation $u_{t} = F$ 
to be strictly pseudo-spherical with associated differential functions $f_{\alpha\beta}$ are the 
conjunction of 

{\em (a)} The functions $f_{\alpha\beta}$ satisfy the constraints:
\begin{equation}
f_{\alpha 1, u_{x^{a}}} = 0 \; , \; \; \; \; \; \; \; \; \; \; 
f_{\alpha 2,u_{x^{k}}} = 0 \; ,  \; \; \; \; \; \; \; \; \; \;
f_{11,u}^{2} + f_{21,u}^{2} + f_{31,u}^{2} ~ \neq ~ 0 \; ,                \label{sve00}
\end{equation}
in which $a \geq 1$ and $\alpha =1,2,3$; and 

{\em (b)} The following identities hold:
\begin{eqnarray}
-f_{11,u}F + \sum_{i=0}^{k-1}u_{x^{i+1}}f_{12,u_{x^{i}}} + f_{21}f_{32} -
f_{31}f_{22} + f_{12,x} - f_{11,t} & = & 0 \; ,                                  \label{sve1} \\
-f_{21,u}F + \sum_{i=0}^{k-1}u_{x^{i+1}}f_{22,u_{x^{i}}} + f_{12}f_{31} -
f_{11}f_{32} + f_{22,x} - f_{21,t} & = & 0 \; , \; \; \; \; \; \;  \mbox{ and} ~ ~ ~
                                                                                 \label{sve2} \\
-f_{31,u}F + \sum_{i=0}^{k-1}u_{x^{i+1}}f_{32,u_{x^{i}}} + f_{21}f_{12} -
f_{11}f_{22} + f_{32,x} - f_{31,t} & = & 0 \; .                                  \label{sve3}
\end{eqnarray}
\end{lemma}  

\section{Hierarchies of equations of pseudo-spherical type}

One considers an affine space equipped with coordinates $(x, t, \tau_{1}, \tau_{2}, 
\dots )$. The independent variable $t$ will be also denoted by $\tau_{0}$ in what follows; 
moreover, in this section and henceforth a {\em differential function}
will be a smooth function depending on a finite number of variables $x, t, \tau_{1}, \tau_{2}, 
\dots$, $u$, and a finite number of $x$--derivatives of $u$. 

Hierarchies of evolution equations $u_{\tau_{i}} = F_{i}$, $i \geq 0$, which describe 
pseudo-spherical surfaces are introduced thus:              

\begin{defn} \label{hi}
Let $u_{\tau_{i}} = F_{i}$, $i \geq 0$, be a countable number of evolutionary equations, in which
$F_{i}$, $i \geq 0$, are differential functions. These equations form a hierarchy of equations 
describing pseudo-spherical surfaces (or, a hierarchy of pseudo-spherical type) if there exist 
differential functions $f_{\alpha\beta}$ and $h_{\alpha i}$ $(\alpha =1,2,3; \beta = 1,2; i 
\geq 1)$, such that for each $n \geq 0$, the  one--forms $\Theta^{[n]}_{\alpha}$ given by 
\begin{equation}
\Theta^{[n]}_{\alpha}= f_{\alpha 1} dx + f_{\alpha 2}dt + \sum_{i=1}^{n}h_{\alpha i}\,d\tau_{i} 
\; ,                          ~ ~ ~ ~ ~ ~ ~ ~  \alpha = 1,2,3 \; ,                \label{ominf}
\end{equation}
satisfy the equations
\begin{equation}
d_{H} \Theta^{[n]}_{1} = \Theta^{[n]}_{3} \wedge \Theta^{[n]}_{2} \; , ~ ~ ~ ~ ~
d_{H} \Theta^{[n]}_{2} = \Theta^{[n]}_{1} \wedge \Theta^{[n]}_{3} \; , ~ ~ ~ ~ ~
d_{H} \Theta^{[n]}_{3} = \Theta^{[n]}_{1} \wedge \Theta^{[n]}_{2} \; .              \label{109}
\end{equation} 
\end{defn}

In (\ref{109}), the exterior derivative $d_{H}\Theta^{[n]}_{\alpha}$ is computed by means of
\begin{equation}
d_{H} g = D_{x}g dx + \sum_{i=0}^{n} D_{\tau_{i}} g\,d\tau_{i} \; ~ ~ ~ \mbox{ and } ~ ~ ~ 
d_{H}(dx) = d_{H}(d\tau_{i})     =0 \;                                             \label{109a}
\end{equation}
for any differential function $g$ and $i \geq 0$, in which the operators $D_{x}$, $D_{\tau_{i}}$,
$i \geq 0$, are given by
\begin{equation}
D_{x} = \frac{\partial}{\partial x} + \sum_{j=0}^{\infty} u_{x^{j+1}}\, 
\frac{\partial}{\partial u_{x^{j}}} \; ,                           ~ ~ ~ ~ \mbox{ and } ~ ~ ~ ~ 
D_{\tau_{i}} = \frac{\partial}{\partial \tau_{i}} + \sum_{j=0}^{\infty}
( D_{x}^{j} F_{i} ) \, \frac{\partial}{\partial u_{x^{j}}} \; .                    \label{109b}
\end{equation}

Two examples of hierarchies of pseudo-spherical type are given in the following section.
Hereafter, if $u_{\tau_{i}} = F_{i}$, $i \geq 0$, is a hierarchy of pseudo-spherical type, 
$u_{t} = F$ will be called the {\em seed equation}, while the equations $u_{\tau_{i}} = F_{i}$, 
$i \geq 1$, will be referred to as the {\em higher} equations of the hierarchy.

\begin{defn}
A local solution of a hierarchy of pseudo-spherical type $u_{\tau_{i}} = F_{i}$, $i \geq 0$, is 
a sequence $\{ u^{[n]}(x,t,\tau_{1},\dots, \tau_{n}) \}_{n\geq 0}$ of smooth functions $u^{[n]}:
V^{[n]} \subset {\bf R}^{n+2} \rightarrow {\bf R}$ such that for each $n \geq 0$, 

$(a)$ $u^{[n]}$ is a solution of the equations $u_{\tau_{i}} = F_{i}$, $i = 0, \dots , n$, and 

$(b)$ $u^{[n+1]}|_{{V}^{[n]}} = u^{[n]}$. 

\end{defn}

The geometrical meaning of Definition \ref{hi} is explored in the next theorem:

\begin{thm} \label{geo}
Let $u_{\tau_{i}} = F_{i}$, $i \geq 0$, be a hierarchy of pseudo-spherical type with
associated one--forms
\begin{equation}
\Theta^{[n]}_{\alpha}= f_{\alpha 1} dx + f_{\alpha 2}dt + \sum_{i=1}^{n} h_{\alpha i}\,d\tau_{i} 
\; ,                                ~ ~ ~ ~ ~ ~ ~ ~  \alpha = 1,2,3 ; ~ n \geq 0 \; , \label{h01}
\end{equation}
and let the sequence of real--valued smooth functions
\[
\{ u^{[n]}(x,t,\tau_{1}, \dots, \tau_{n}) \}_{n\geq 0}
\] 
be an arbitrary solution of the hierarchy $u_{\tau_{i}} = F_{i}$. For each $n \geq 0$, let
$V^{[n]}$ be (an open subset of) the space of independent variables 
$(x,t,\tau_{1},\dots,\tau_{n})$ contained in the domain of the function $u^{[n]}$, and assume that
$V^{[n]}$ is equipped with a flat pseudo-Riemannian metric of index $s$.
Assume further that 
\[
\Theta^{[n]}_{1} \wedge \Theta^{[n]}_{2} \, ( u^{[n]}(x,t,\tau_{1},\dots,\tau_{n}) ) \neq 0
\]
on $V^{[n]}$. Then, if the index of $V^{[n]}$ is $s = n$, there exists a 
two--dimensional immersed Riemannian submanifold $S$ of $V^{[n]}$ such that 

(a) The one--forms 
\begin{equation}
\Theta^{[n]}_{1}( u^{[n]}(x,t,\tau_{1},\dots,\tau_{n}) ) \, , \; \; \; \mbox{ and } \; \; \; 
\Theta^{[n]}_{2} \, ( u^{[n]}(x,t,\tau_{1},\dots,\tau_{n}) ) \;             \label{le0}
\end{equation}
are a moving orthonormal coframe on $S$ and the one--form 
$\Theta^{[n]}_{3}( u^{[n]}(x,t,\tau_{1},\dots,\tau_{n}) )$ is the corresponding Levi--Civita 
connection one--form.

(b) The induced metric of $S$ has constant curvature $-1$.

(c) The normal bundle of the immersion is flat, that is, the normal curvature of $S$ is 
    identically zero.
\end{thm}

\begin{proof}
The following range of indices will be used in what follows:
\[
1 \leq A, B \leq n + 2 \; ; \; \; \; \; \; 1 \leq i,j \leq 2 \; ; \; \; \; \; \;
3 \leq \alpha , \beta \leq n+2 \; .
\]
Set $\sigma_{A} = 1$, except for $s$ indices between $3$ and $n+2$ for which 
$\sigma_{A} = -1$. The structure equations of a two--dimensional manifold $S$ 
immersed in the flat space $V^{[n]}$ are \cite{T}
\begin{equation}
d \, \omega_{1} = \omega_{2} \wedge \omega_{21} \; , \; \; \; \; \; 
d \, \omega_{2} = \omega_{1} \wedge \omega_{12} \; ,                    \label{le1}
\end{equation}
together with the Gauss equation
\begin{equation}
 d \, \omega_{12} = \sum_{\alpha = 3}^{n+2} \sigma_{\alpha} \, \omega_{1 \alpha} \wedge 
                    \omega_{\alpha 2} \; ,                                           \label{le3}
\end{equation}
the Codazzi equations
\begin{equation}
d \, \omega_{i \alpha} = \sum_{j=1}^{2} \omega_{ij} \wedge \omega_{j \alpha} + 
\sum_{\beta = 3}^{n+2} \sigma_{\beta} \, \omega_{i \beta} \wedge \omega_{\beta \alpha} \; ,
									         \label{le4}
\end{equation}
and the Ricci equations 
\begin{equation}
d \, \omega_{\alpha \beta} = \sum_{\gamma = 3}^{n+2} \sigma_{\gamma} \, \omega_{\alpha \gamma} 
                             \wedge \omega_{\gamma \beta} + \Omega_{\alpha \beta} \; ,
									         \label{le5}
\end{equation}
in which
\begin{equation}
\Omega_{\alpha \beta} = \omega_{\alpha 1} \wedge \omega_{1 \beta} +             \label{le6}
                        \omega_{\alpha 2} \wedge \omega_{2 \beta}
\end{equation}
is the normal curvature of $S$. Moreover, the connection one--forms $\omega_{AB}$ satisfy
\begin{equation}
\omega_{1} \wedge \omega_{1 \alpha} + \omega_{2} \wedge \omega_{2 \alpha} = 0 \; , \; \; \; \;
\mbox{ and } \; \; \; \; \omega_{AB} + \omega_{BA} = 0      \; .                      \label{le2}
\end{equation}
Conversely, such an immersed surface exists if there exist one--forms $\omega_{i}$ and 
$\omega_{AB}$ satisfying Equations (\ref{le1})--(\ref{le2}).

Now, Definition \ref{hi} implies that Equations (\ref{le1}) are satisfied if one sets 
\begin{eqnarray}
\omega_{1} = \Theta^{[n]}_{1}( u^{[n]}(x,t,\tau_{1},\dots,\tau_{n}) ) \; , & & 
\omega_{2} = \Theta^{[n]}_{2}( u^{[n]}(x,t,\tau_{1},\dots,\tau_{n}) ) \; ,     \label{le6.5} \\
&  & \omega_{12} = \Theta^{[n]}_{3}( u^{[n]}(x,t,\tau_{1},\dots,\tau_{n}) ) \; .
\end{eqnarray}
Moreover, since Equations (\ref{le2}) imply that
\begin{equation}
\omega_{i \alpha} = h^{\alpha}_{i1} \, \omega_{1} + h^{\alpha}_{i 2} \, \omega_{2} \; , \; \; 
\mbox{ and } \; \; h^{\alpha}_{ij} = h^{\alpha}_{ji} \; ,                         \label{le7}
\end{equation}
one can take
\begin{equation}
\omega_{1 \alpha} = \frac{1}{\sqrt{n}} \, \omega_{1} \; , \; \; \; \; \; \mbox{ and } 
\; \; \; \; \; 
\omega_{2 \alpha} = \frac{1}{\sqrt{n}} \, \omega_{2} \; ,                        \label{le8}
\end{equation}
and define one--forms $\omega_{\alpha \beta}$ simply by
\begin{equation}
\omega_{\alpha \beta} = 0 \; .                                                  \label{le9}
\end{equation}
Equation (\ref{le9}) implies that the Ricci equations (\ref{le5}) become 
$\Omega_{\alpha\beta} = 0$, and these equations are identically satisfied because of (\ref{le8}).
Also, (\ref{le8}) and (\ref{le9}) imply that the Codazzi equations (\ref{le4}) read 
\[
d\,\omega_{1} = \omega_{12} \wedge \omega_{2} \; , \; \; \; \; \;
d\,\omega_{2} = \omega_{21} \wedge \omega_{1} \; ,
\]
which are precisely Equations (\ref{le1}). Finally, the Gauss equation (\ref{le3}) becomes
\[
d\,\omega_{12} = \sum_{\alpha = 3}^{n+2} \sigma_{\alpha} \, \frac{1}{n} \, 
                 \omega_{1} \wedge (- \omega_{2}) = - \frac{1}{n} \, 
                 (\sum_{\alpha = 3}^{n+2} \sigma_{\alpha}) \, \omega_{1} \wedge \omega_{2} \; ,
\]
and this equation holds if one takes 
\[
\sigma_{\alpha} = -1 \; 
\]
for all $\alpha$, so that the index of the pseudo-Riemannian manifold $V^{[n]}$ is $s = n$. 
Thus, Equations (\ref{le6.5})--(\ref{le9}) determine an immersed Riemannian surface $S$ 
of constant curvature $-1$ described by the moving coframe (\ref{le0}), and such that 
the normal bundle of the immersion is flat.
\end{proof}

Theorem \ref{geo} is a higher dimensional analog of a straightforward immersion result for single
equations of pseudo-spherical type appearing in \cite{175,19}. An obvious problem that now 
presents itself is whether there exists a relationship between the hierarchies of 
pseudo-spherical type considered in this paper and the intrinsic generalizations of 
two--dimensional integrable equations studied by K. Tenenblat and her co--workers \cite{T}, 
which are related to immersions of manifolds of constant curvature into higher dimensional 
space forms. This issue will be considered in a separate work. 

The following lemma allows one to prove that each equation $u_{\tau_{i}} = F_{i}$
of a hierarchy of pseudo-spherical type describes pseudo-spherical surfaces.

\begin{lemma}   \label{lemma1}
Let $u_{\tau_{i}} = F_{i}$, $i \geq 0$, be a hierarchy of pseudo-spherical type with
associated one--forms
\begin{equation}
\Theta^{[n]}_{\alpha}= f_{\alpha 1} dx + f_{\alpha 2}dt + \sum_{i=1}^{n} h_{\alpha i}\,d\tau_{i} 
\; ,                                ~ ~ ~ ~ ~ ~ ~ ~  \alpha = 1,2,3 ; ~ n \geq 0 \; . \label{h11}
\end{equation}
Then, the following four sets of equations hold:
\begin{eqnarray}
 &
\left\{
\begin{array}{ccc}
-D_{t}{f}_{11} + D_{x}{f}_{12} & = & {f}_{31}{f}_{22} - {f}_{32}{f}_{21}   \\
-D_{t}{f}_{21} + D_{x}{f}_{22} & = &  {f}_{11} {f}_{32} - {f}_{12} {f}_{31}   \\
-D_{t}{f}_{31} + D_{x}{f}_{32} & = &  {f}_{11} {f}_{22} - {f}_{12} {f}_{21} 
\end{array}
\right.
 &   \label{h31} \\
 &
\left\{
\begin{array}{ccc}
- D_{\tau_{i}}{f}_{11} +  D_{x}{h}_{1i} & = &  {f}_{31} {h}_{2i} -  {h}_{3i} {f}_{21}   \\
- D_{\tau_{i}}{f}_{21} +  D_{x}{h}_{2i} & = &  {f}_{11} {h}_{3i} -  {h}_{1i} {f}_{31}  \\
- D_{\tau_{i}}{f}_{31} +  D_{x}{h}_{3i} & = &  {f}_{11} {h}_{2i} -  {h}_{1i} {f}_{21}
\end{array}
\right.
 &  \label{h41}   \\ 
 &
\left\{      \begin{array}{ccc}
- D_{\tau_{i}}{f}_{12} + D_{t}{h}_{1i} & = &  {f}_{32} {h}_{2i} -  {h}_{3i} {f}_{22}   \\
- D_{\tau_{i}}{f}_{22} + D_{t}{h}_{2i} & = &  {f}_{12} {h}_{3i} -  {h}_{1i} {f}_{32}  \\
- D_{\tau_{i}}{f}_{32} + D_{t}{h}_{3i} & = &  {f}_{12} {h}_{2i} -  {h}_{1i} {f}_{22} 
\end{array}
\right.
  &  \label{h51} \\
  &
\left\{
\begin{array}{ccc}
- D_{\tau_{j}}{h}_{1i} +  D_{\tau_{i}}{h}_{1j} & = &  {h}_{3i} {h}_{2j} -  {h}_{3j} {h}_{2i}   \\
- D_{\tau_{j}}{h}_{2i} +  D_{\tau_{i}}{h}_{2j} & = &  {h}_{1i} {h}_{3j} -  {h}_{1j} {h}_{3i}   \\
- D_{\tau_{j}}{h}_{3i} +  D_{\tau_{i}}{h}_{3j} & = &  {h}_{1i} {h}_{2j} -  {h}_{1j} {h}_{2i} 
\end{array}
\right.
  &  \label{h61}
\end{eqnarray}   
in which $i,j \geq 1$. Conversely, if $u_{\tau_{i}} = F_{i}$, $i \geq 0$, is a sequence of 
evolution equations, the operators $D_{x}$ and $D_{\tau_{i}}$ are defined by means of 
$(\ref{109b})$, and there exist differential functions $f_{\alpha\beta}$, $h_{\alpha i}$ 
$(\alpha = 1,2,3$; $\beta = 1,2$; $i \geq 1)$ such that $(\ref{h31})$--$(\ref{h61})$ are 
satisfied, then $u_{\tau_{i}} = F_{i}$, $i \geq 0$, is a hierarchy of pseudo-spherical type 
with associated one--forms $(\ref{h11})$.
\end{lemma}   

The proof of Lemma \ref{lemma1} is elementary and will be omitted. One now has:

\begin{prop} \label{prop4}
Let $u_{\tau_{i}} = F_{i}$, $i \geq 0$, be a hierarchy of pseudo-spherical type with associated 
one--forms $\Theta^{[n]}_{\alpha}= f_{\alpha 1} dx + f_{\alpha 2}dt + 
\sum_{i=1}^{n}h_{\alpha i}d\tau_{i}$, $\alpha = 1,2,3$, $n \geq 0$. Then, the equations 
$u_{\tau_{i}} = F_{i}$, $i \geq 0$, describe pseudo-spherical surfaces with associated one--forms
$\omega^{\alpha}_{0} = f_{\alpha 1}dx +f_{\alpha 2}dt$ if $i = 0$, and $\omega^{\alpha}_{i} = 
f_{\alpha 1}dx + h_{\alpha i} d\tau_{i}$, if $i \geq 1$. 
\end{prop}

\begin{remark}
If $u_{\tau_{i}} = F_{i}$, $i \geq 0$ is a hierarchy of pseudo-spherical type, 
Equations (\ref{linear1}), (\ref{linear2}) and (\ref{109}) imply that for each $n \geq 0$, 
the trivial bundle $V^{n+2} \times SL(2, {\bf R})$ can be equipped ---whenever 
$\{ u^{[n]}(x,t,\tau_{1},\dots, \tau_{n}) \}_{n\geq 0}$ is a solution to the hierarchy 
$u_{\tau_{i}} = F_{i}$--- with a flat $sl(2,{\bf R})$--valued connection, in analogy with the
single equation case. Moreover, Proposition \ref{prop4} implies that for each $i \geq 0$, the 
pull--back of this connection to the submanifold of coordinates $(x,\tau_{i})$ gives a flat 
$sl(2,{\bf R})$--valued connection form associated with the equation $u_{\tau_{i}} = F_{i}$.
\end{remark}

In the remainder of this section, it will be proven that Definition \ref{hi} encodes the main 
properties of standard hierarchies of ``integrable'' equations alluded to in the Introduction. 

\begin{prop}
Let $u_{\tau_{i}} = F_{i}$, $i \geq 0$, be a hierarchy of pseudo-spherical type. The
equations $u_{\tau_{i}} = F_{i}$, $i \geq 0$, are the integrability condition of linear problems
with a common space part, that is, there exist matrix--valued functions $X$ and $T_{i}$ such 
that for each $i \geq 0$, $u_{\tau_{i}} = F_{i}$ is the integrability condition of the linear 
problem
\[
v_{x} = X\,v \; , \; \; \; \; \; \; \; \; \; \; v_{\tau_{i}} = T_{i}\,v \; .
\]
\end{prop}
\begin{proof}
This result follows from Proposition \ref{prop4} and Equations (\ref{linear1})--(\ref{zcc}).
\end{proof}

Next, one would like to show that hierarchies of pseudo-spherical type generate hierarchies of 
pairwise commuting flows. In order to avoid some technicalities ---mentioned in Remark 
\ref{gauge} at the end of this section--- this property is proven 
below in the special case of hierarchies of {\em strictly} pseudo-spherical
type, which are now defined:
              
Consider a countable number of evolution equations $u_{\tau_{i}} = F_{i}$, of order $k_{i}$, 
$i \geq 0$. For each $n \geq 0$, let $J^{(n)}$ be a manifold with coordinates $(x,t,\tau_{1}, 
\dots , \tau_{n}, u, u_{x}, \dots, u_{x^{M(n)}})$, where $M(n)$ is the maximum of the orders 
$k_{j}$, $0 \leq j \leq n$, and ---in analogy with the single equation case--- let $I^{(n)}$ be 
the differential ideal generated by the two--forms
\begin{eqnarray}
du \wedge dx ~ + ~ F_{0}\,dx \wedge dt & + & \sum_{i=1}^{n} F_{i}\,dx \wedge d\tau_{i} \; , \\
du_{x^{k}} \wedge dt ~ - ~ u_{x^{k+1}} dx \wedge dt & - & 
\sum_{j=1}^{n} D_{x}^{k} F_{j}\,d\tau_{j} \wedge dt, ~ ~ ~ 0 \leq k \leq k_{0} - 1 \; , 
                                                                              \; \; \; \; \; \\
\mbox{and} \hspace{55mm} & & ~ \nonumber \\
du_{x^{k}} \wedge d\tau_{j} ~ - ~ u_{x^{k+1}} dx \wedge d\tau_{j} & - & D_{x}^{k} F_{0}\,dt 
\wedge d\tau_{j} ~ - ~ \sum_{l=1}^{n} D_{x}^{k}F_{l}\,d\tau_{l}\wedge d \tau_{j} \; , 
                                                                                  \; \; \; \; \; 
\end{eqnarray}
in which for each $j = 1, \dots, n$, $0 \leq k \leq k_{j} - 1$.

\begin{defn} \label{str}
A countable collection of evolution equations $u_{\tau_{i}} = F_{i}$, $i \geq 0$, is a hierarchy
of strictly pseudo-spherical type if there exist differential functions $f_{\alpha\beta}$, 
$\alpha =1,2,3$; $\beta = 1,2$, and $h_{\alpha i}$, $i \geq 1$, such that for every $n \geq 0$, 
the two--forms
\begin{equation}
\Omega_{1}^{[n]} = d \Theta^{[n]}_{1} - \Theta^{[n]}_{3} \wedge \Theta^{[n]}_{2} \; , \; \; \; \;
\Omega_{2}^{[n]} = d \Theta^{[n]}_{2} - \Theta^{[n]}_{1} \wedge \Theta^{[n]}_{3} \; , \; \; \; \;
\Omega_{3}^{[n]} = d \Theta^{[n]}_{3} - \Theta^{[n]}_{1} \wedge \Theta^{[n]}_{2} \; , \;  
\end{equation}
in which 
\begin{equation}
\Theta^{[n]}_{\alpha}= f_{\alpha 1} dx + f_{\alpha 2}dt + \sum_{i=1}^{n}h_{\alpha i}d\tau_{i} 
\; ,                          ~ ~ ~ ~ ~ ~ ~ ~  \alpha = 1,2,3,                  \label{ostrict}
\end{equation}
 generate the ideal $I^{(n)}$.
\end{defn}

Instead of Lemma \ref{lemma0} on  a single strictly pseudo-spherical equation, one now has:

\begin{prop}   \label{hstrict}
Consider a sequence of equations $u_{\tau_{i}} = F_{i}$ of order $k_{i}$, $i \geq 0$, and for 
each $n \geq 0$ let $M(n)$ be the maximum of the orders $k_{j}$, $0 \leq j \leq n$. Necessary and
sufficient conditions for the equations $u_{\tau_{i}} = F_{i}$, $i \geq 0$, to form a hierarchy 
of strictly pseudo-spherical type with associated functions $f_{\alpha\beta}$ and $h_{\alpha i}$
$(\alpha =1,2,3; \beta = 1,2; i \geq 1)$ are the conjunction of 

{\rm (a)} The functions $f_{\alpha 1}$ satisfy  
          $f_{11,u}^{2} + f_{21,u}^{2} + f_{31,u}^{2} \neq 0$ ; 

{\rm (b)} For each $n \geq 0$, the functions $f_{\alpha\beta}$ and $h_{\alpha i}$ satisfy 
\begin{eqnarray*}
f_{\alpha 1,u_{x^{k}}} = 0 \; , \; \; 1 \leq k \leq M(n) \; ; \; \; 
f_{\alpha 2,u_{x^{k}}} = 0 \; , & & k_{0} \leq k \leq M(n) \; ; \\
h_{\alpha i,u_{x^{k}}} = 0 \; , & &  k_{i} \leq k \leq M(n) \; ; 
\end{eqnarray*}
and 

{\rm (c)} The functions $F_{i}$, $f_{\alpha\beta}$, and $h_{\alpha i}$ satisfy the identities
\begin{eqnarray}
&
\left\{
\begin{array}{ccc}
-(f_{11,t} + f_{11,u}F_{0}) + D_{x}{f}_{12} & = & {f}_{31} {f}_{22} - {f}_{32} {f}_{21}  \\
-(f_{21,t} + f_{21,u}F_{0}) + D_{x}{f}_{22} & = & {f}_{11} {f}_{32} - {f}_{12} {f}_{31}  \\
-(f_{31,t} + f_{31,u}F_{0}) + D_{x}{f}_{32} & = & {f}_{11} {f}_{22} - {f}_{12} {f}_{21}
\end{array}
\right.
 &   \label{h311} \\   
 &
\left\{
\begin{array}{ccc}
-(f_{11,\tau_{i}} + f_{11,u}F_{i}) + D_{x}{h}_{1i} & = &  {f}_{31}{h}_{2i} -  {h}_{3i}{f}_{21} \\
-(f_{21,\tau_{i}} + f_{21,u}F_{i}) + D_{x}{h}_{2i} & = &  {f}_{11}{h}_{3i} -  {h}_{1i}{f}_{31} \\
-(f_{31,\tau_{i}} + f_{31,u}F_{i}) + D_{x}{h}_{3i} & = &  {f}_{11}{h}_{2i} -  {h}_{1i}{f}_{21} 
\end{array}
\right.      
&        \label{h411}    \\ 
&
\left\{
\begin{array}{ccc}
- D_{\tau_{i}}{f}_{12} + D_{t}{h}_{1i} & = &  {f}_{32} {h}_{2i} -  {h}_{3i} {f}_{22}   \\
- D_{\tau_{i}}{f}_{22} + D_{t}{h}_{2i} & = &  {f}_{12} {h}_{3i} -  {h}_{1i} {f}_{32}  \\
- D_{\tau_{i}}{f}_{32} + D_{t}{h}_{3i} & = &  {f}_{12} {h}_{2i} -  {h}_{1i} {f}_{22}
\end{array}
\right.
  &      \label{h511} \\
  &
\left\{
\begin{array}{ccc}
- D_{\tau_{j}}{h}_{1i} +  D_{\tau_{i}}{h}_{1j} & = &  {h}_{3i} {h}_{2j} -  {h}_{3j} {h}_{2i}   \\
- D_{\tau_{j}}{h}_{2i} +  D_{\tau_{i}}{h}_{2j} & = &  {h}_{1i} {h}_{3j} -  {h}_{1j} {h}_{3i}   \\
- D_{\tau_{j}}{h}_{3i} +  D_{\tau_{i}}{h}_{3j} & = &  {h}_{1i} {h}_{2j} -  {h}_{1j} {h}_{2i} 
\end{array}
\right.       \label{h611}
\end{eqnarray}
for all $i,j \geq 0$.
\end{prop} 
\begin{proof}
Expand the two--form $\Omega_{1}^{[n]}$, and consider the structure equation 
$\Omega_{1}^{[n]} = 0$. Definition \ref{str} implies that
\begin{eqnarray}
du \wedge dx ~ + ~ F_{0} dx \wedge dt & + & \sum_{i=1}^{n} F_{i} \,dx \wedge d\tau_{i} ~ = ~ 0 
                                                                                         \; , \\
du_{x^{k}} \wedge dt ~ - ~ u_{x^{k+1}} dx \wedge dt & - & \sum_{j=1}^{n} D_{x}^{k}F_{j} \,  
d\tau_{j} \wedge dt ~ = ~ 0 \; , ~ ~ ~ 0 \leq k \leq k_{0} - 1 \; , 
\; \; \; \; \; \; \; \;  \\ 
\mbox{and} \hspace{55mm} & & ~ \nonumber \\
du_{x^{k}} \wedge d\tau_{j} ~ - ~ u_{x^{k+1}} dx \wedge d\tau_{j} & - & D_{x}^{k} F_{0} \,dt 
\wedge d\tau_{j} ~ - ~ \sum_{l=1}^{n} D_{x}^{k}F_{l} \,d\tau_{l} \wedge d \tau_{j} ~ = 0 \; , 
                                           \; \; \; \; \; \; \; 
\end{eqnarray}
in which for each $j = 1, \dots, n$, $0 \leq k \leq k_{j} - 1$. Substituting these equations into
$\Omega_{1}^{[n]} = 0$, and collecting terms, one finds the identities
\begin{eqnarray}
-f_{11,t} - f_{11,u}F_{0} + D_{x}f_{12} - f_{31}f_{22} + f_{32}f_{21} & = & 0 \; , \\
-f_{11,\tau_{i}} - f_{11,u}F_{i} + D_{x}h_{1i} - f_{31}h_{2i} + h_{3i}f_{21} & = & 0 \; , \\
-D_{\tau_{i}}f_{12} + D_{t}h_{1i} - f_{32}h_{2i} + h_{3i}f_{22} & = & 0 \; , \\
-D_{\tau_{j}}h_{1i} + D_{\tau_{i}}h_{1j} - h_{3i}h_{2j} + h_{3j}h_{2i} & = & 0 \; ,
\end{eqnarray}
and 
\begin{eqnarray}
f_{1 1,u_{x^{k}}} & = & 0, ~ ~ 1 \leq k \leq M(n) \; ;   \\
f_{1 2,u_{x^{k}}} & = & 0, ~ ~ k_{0} \leq k \leq M(n) \; ; \\
h_{1 i,u_{x^{k}}} & = & 0, ~ ~ k_{i} \leq k \leq M(n) \; .
\end{eqnarray}
The other equations appearing in (\ref{h31})--(\ref{h61}) are obtained from the  
equations $\Omega_{2}^{[n]} = \Omega_{3}^{[n]} = 0$. Finally, the constraint $f_{11,u}^{2} + 
f_{21,u}^{2} + f_{31,u}^{2} \neq 0$ holds, for otherwise the equations $u_{\tau_{i}}=F_{i}$ 
would not be necessary and sufficient for the structure equations $\Omega^{[n]}_{\alpha} = 0$ 
to be satisfied. 
\end{proof}

\begin{cor}
Assume that the equations $u_{\tau_{i}} = F_{i}$, $i \geq 0$, form a hierarchy 
of strictly pseudo-spherical type with associated functions $f_{\alpha\beta}$ and 
$h_{\alpha i}$ $(\alpha =1,2,3; \beta = 1,2; i \geq 1)$. Then, each equation 
$u_{\tau_{i}} = F_{i}$ is strictly pseudo-spherical with associated
one--forms $\omega^{\alpha}_{0} = f_{\alpha 1}dx +f_{\alpha 2}dt$
$($if $i = 0)$ and $\omega^{\alpha}_{i} = f_{\alpha 1}dx + h_{\alpha i} d\tau_{i}$ $($if 
$i \geq 1)$ in which $\alpha =1,2,3$.
\end{cor}

In order to prove that a hierarchy $u_{\tau_{i}} = F_{i}$ of strictly pseudo-spherical 
type generates a family of pairwise commuting flows, one needs to show that the function 
$F_{j}$ is a {\em generalized symmetry} of the equation $u_{\tau_{i}} = F_{i}$ for all $i,j 
\geq 0$:

\begin{defn}
A differential function $G$ is a generalized symmetry of an evolution equation 
$u_{t}=F(x,t,u,\dots,u_{x^{n}})$ if for any local solution $u(x,t)$, the function $u(x,t) + 
\tau G(u(x,t))$ satisfies $u_{t}=F$ to first order in $\tau$.
\end{defn}   
 
Thus, $G$ is a generalized symmetry of $u_{t}=F$ if and only if the equation $D_{t} G = F_{\ast}
G$, in which for any $f(x,t,u, \dots , u_{x^{k}})$, $f_{\ast}$ is the {\em formal linearization} 
of $f$ given by \cite{16}
\begin{equation}
f_{\ast} = \sum_{i=0}^{k} \frac{\partial f}{\partial u_{x^{i}}} \, D_{x}^{i} \; ,  \label{forli}
\end{equation}
holds identically once all the derivatives with respect to $t$ appearing in it have been 
replaced by means of $u_{t}=F$. Equivalently \cite{16}, an evolution equation $u_{t} = G$ 
is a generalized symmetry of $u_{t}=F$ if, at least formally, their flows commute, that is, if
\begin{equation}
\frac{\partial G}{\partial t} + F_{\ast}G - G_{\ast}F = 0 \label{sym}
\end{equation}
whenever $u(x,t)$ is a solution of $u_{t}=F$.

The following characterization of generalized symmetries of strictly pseudo-spherical 
evolution equations holds \cite{19}:

\begin{lemma} \label{sim}
Let $u_{t} = F(x,t,u, \dots, u_{x^{m}})$ be a strictly pseudo-spherical evolution equation with 
associated one-forms $\omega^{\alpha} = f_{\alpha 1}dx + f_{\alpha 2}dt$, $\alpha = 1,2,3$. 
Let $G$ be a differential function, and let $u(x,t)$ be a local solution of $u_{t} = F$. 
Consider the deformed one--forms 
\begin{equation} \label{defm}
{\omega}^{\alpha}(u(x,t)) + \tau {\Lambda}_{\alpha}(u(x,t)) \; ,
\end{equation} 
in which
\begin{equation}
{\Lambda}_{\alpha}(u(x,t)) = f_{\alpha 1,u}(u(x,t))\,G(u(x,t)) \,dx + \sum_{i=0}^{m-1}
f_{\alpha 2,u_{x^{i}}}(u(x,t))\,\frac{\partial^{i}G(u(x,t))}{\partial x^{i}}\,dt \; .
								\label{defor}
\end{equation}
Then, the deformed one--forms ${\omega}^{\alpha}(u(x,t)) + \tau 
{\Lambda}_{\alpha}(u(x,t))$ satisfy the structure equations of a pseudo-spherical surface up to 
terms of order $\tau^{2}$ if and only if $G$ is a generalized symmetry of the equation 
$u_{t} = F$. 
\end{lemma}

The final theorem of this section reads as follows:

\begin{thm}  \label{sym0}
Assume that $u_{\tau_{i}} = F_{i}$, $i \geq 0$, is a hierarchy of strictly pseudo-spherical type
with associated one--forms
\begin{equation}
\Theta^{[n]}_{\alpha} = f_{\alpha 1}dx + f_{\alpha 2}dt + \sum_{i=1}^{n} h_{\alpha i}
d \tau_{i} \; , ~ ~ ~ \alpha =1,2,3; ~ n \geq 0    \;     .      \label{oinf}
\end{equation}
The function $F_{j}$ is a generalized symmetry of the equation $u_{\tau_{i}} = F_{i}$, for all 
$i, j \geq 0$.
\end{thm}       
\begin{proof} 
Since $u_{\tau_{i}} = F_{i}$, $i \geq 0$ is a hierarchy of strictly pseudo-spherical type, 
the equations $u_{t} = F$ and $u_{\tau_{i}} = F_{i}$ are strictly pseudo-spherical 
with associated one forms $\omega_{0}^{\alpha} = f_{\alpha 1} dx + f_{\alpha 2}dt$ and 
$\omega^{\alpha}_{i} = f_{\alpha 1}dx + h_{\alpha i}d\tau_{i}$ respectively, and therefore 
one can apply the last lemma. 

One checks first that the equations $u_{\tau_{i}} = F_{i}$ determine generalized symmetries of 
the seed equation $u_{t} = F$. Consider the deformations of the one--forms 
$\omega_{0}^{\alpha}$  induced by
\[
u \mapsto u + \tau F_{i} \; ,
\]
and set 
\[
\Lambda_{0}^{\alpha} = f_{\alpha 1,\tau_{i}} dx + f_{\alpha 2,\tau_{i}} dt \; ,
\]
$\alpha = 1,2,3$. These one--forms are of the type (\ref{defor}) if pulled--back by solutions
$u(x,t)$ of $u_{t} = F$ because of Proposition \ref{hstrict}. A straightforward computation shows
that the deformed one--forms $\omega_{0}^{\alpha} + \tau_{i}\Lambda_{0}^{\alpha}$ 
describe pseudo-spherical surfaces to first order in $\tau_{i}$ if and only if the equations
\begin{eqnarray}
d {\Lambda}_{1} & = & {\omega}^{3} \wedge {\Lambda}_{2} +
{\Lambda}_{3} \wedge {\omega}^{2} \, ,                           \label{lse4}\\     
d {\Lambda}_{2} & = & {\omega}^{1} \wedge {\Lambda}_{3} +
{\Lambda}_{1} \wedge {\omega}^{3} \, ,  ~ ~ ~ ~ ~ ~ ~ ~ \mbox{ and}
       \label{lse5} \\
d {\Lambda}_{3} & = & {\omega}^{1} \wedge {\Lambda}_{2} +
{\Lambda}_{1} \wedge {\omega}^{2} \, ,                        \label{lse6}
\end{eqnarray}
hold whenever $u(x,t)$ is a solution of the equation $u_{t} = F$. Now, Equation (\ref{lse4}), for
instance, holds if and only if
\begin{equation}
-f_{11,\tau_{i}t} + f_{12,\tau_{i}x} = f_{31}f_{22,\tau_{i}} - f_{32}f_{21,\tau_{i}}
+ f_{31,\tau_{i}}f_{22} - f_{21}f_{32,\tau_{i}} \; .      \label{lse7}
\end{equation}
Because of Equations (\ref{h411}) and (\ref{h511}), Equation (\ref{lse7}) is equivalent to
\[
-h_{2i}(-f_{31,t} + f_{32,x}) + h_{3i}(-f_{21,t} + f_{22,x}) = -h_{2i}(f_{22}f_{11} -
f_{21}f_{12}) + h_{3i}(f_{32}f_{11} - f_{31}f_{12}) \; ,
\]
an equation which does hold whenever $u(x,t)$ is a solution of $u_{t}=F$, since this
equation describes pseudo-spherical surfaces with associated one--forms
$\omega_{0}^{\alpha}$. Equations (\ref{lse5}) and (\ref{lse6}) are treated in analogous
fashion.
 
One now checks that the equations $u_{\tau_{j}} = F_{j}$ determine generalized symmetries 
of the equations $u_{\tau_{i}} = F_{i}$, $i \neq j$. Recall once more that the
equations $u_{\tau_{i}} = F_{i}$ describe pseudo-spherical surfaces with associated one--forms 
$\omega^{\alpha}_{i} = f_{\alpha 1}dx + h_{\alpha i}d\tau_{i}$. Deform these one--forms by means
of
\[
u \mapsto u + \tau_{j} F_{j} \; .
\]
As before, the one--forms
\[
\Lambda_{i}^{\alpha} = f_{\alpha 1,\tau_{j}} dx + h_{\alpha i,\tau_{j}} d\tau_{i} 
\]
are of the type (\ref{defor}) if pulled back by solutions $u(x,\tau_{i})$ of $u_{\tau_{i}} = 
F_{i}$ because of Proposition \ref{hstrict}. One then needs to prove that the deformed one--forms
$\omega_{0}^{\alpha} + \tau_{i}\Lambda_{i}^{\alpha}$ describe pseudo-spherical surfaces to first 
order in $\tau_{j}$, or, in other words, that the equations
\begin{eqnarray}
d {\Lambda}_{i}^{1} & = & {\omega}^{3}_{i} \wedge {\Lambda}_{i}^{2} +
{\Lambda}_{i}^{3} \wedge {\omega}^{2}_{i} \, ,
     \label{lse8}\\ 
d {\Lambda}_{i}^{2} & = & {\omega}^{1}_{i} \wedge {\Lambda}_{i}^{3} +
{\Lambda}_{i}^{1} \wedge {\omega}^{3}_{i} \, ,  ~ ~ ~ ~ ~ ~ ~ ~ \mbox{ and}  \label{lse9}\\
d {\Lambda}^{3}_{i} & = & {\omega}^{1}_{i} \wedge {\Lambda}_{i}^{2} +
{\Lambda}_{i}^{1} \wedge {\omega}^{2}_{i} \, ,
     \label{lse10}
\end{eqnarray}
are satisfied whenever $u(x,\tau_{i})$ is a solution of the equation $u_{\tau_{i}} = F_{i}$. 
The proof goes as before, using this time Equations (\ref{h511}) and (\ref{h611}).             
\end{proof}

\begin{remark}   \label{gauge}
It is possible to generalize Theorem \ref{sym0} to the case of arbitrary hierarchies of
pseudo-spherical type, but some extra technical difficulties appear. Essentially, one needs
to consider generalized symmetries of arbitrary PSS equations, and state an analog of Lemma
\ref{sim} for them. Such a result can be proven by taking into account the gauge invariance
of the of the theory of equations describing pseudo-spherical surfaces, as in \cite{re}. 
\end{remark}

\section{Correspondence results}
     
The geometric correspondence theorems between solutions of PSS equations which have been recently
obtained by Kamran and Tenenblat \cite{12} and Reyes \cite{re} are based on two fundamental 
facts: first, each generic solution $u(x,t)$ of a PSS equation determines a pseudo-spherical
metric on (an open subset of) the graph of $u(x,t)$; second, surfaces of constant Gaussian
curvature are locally isometric. The authors of \cite{12} noted that these two observations
allow one to establish correspondences between solutions of PSS equations, by explicitly 
constructing these isometries. The transformations one obtains are very different from 
classical B\"{a}cklund transformations \cite{BA,T} for ``soliton equations''. For example, the 
new transformations require explicit changes of independent variables; they appear to be 
unrelated to symmetry considerations (see Chern and Tenenblat \cite{6}, Tenenblat \cite{T}, and 
Beals, Rabelo, and Tenenblat \cite{brt} for relations between symmetries and standard 
B\"{a}cklund correspondences); finally, they are not restricted to transforming solutions of a 
same equation, or even equations of the same order. 

The following three results appear in \cite{re}:

\begin{thm}  \label{olya1}
Let $\Xi(x,t,u, \dots ) = 0$ and $\widehat{\Xi} (\widehat{x},\widehat{t}, \widehat{u}, \dots ) 
= 0$ be two PSS equations with associated one--forms $\omega^{\alpha} = f_{\alpha 1}dx + 
f_{\alpha 2}dt$ and $\widehat{\omega}^{i} = \widehat{f}_{\alpha 1}d\widehat{x} + 
\widehat{f}_{\alpha 2}d\widehat{t}$, $\alpha=1,2,3$, respectively, and assume that these 
one--forms satisfy $\omega^{2} \wedge \omega^{3} \not \equiv 0$, and $\widehat{\omega}^{2} 
\wedge \widehat{\omega}^{3} \not \equiv 0$. Then, for any $I$--generic solutions $u(x,t)$ of 
$\Xi = 0$ and $\widehat{u}(\widehat{x},\widehat{t}\,)$ of $\widehat{\Xi} = 0$, there exist a 
local diffeomorphism $\Psi : V \rightarrow \widehat{V}$, $\Psi(x,t) = (\gamma (x,t), 
\delta(x,t))$ in which $V$ and $\widehat{V}$ are open subsets of the domains of $u(x,t)$ and 
$\widehat{u}(\widehat{x},\widehat{t}\,)$ respectively, and a smooth function $\nu : V 
\rightarrow {\bf R}$, such that $\omega^{\alpha}(u(x,t))$ and 
$\widehat{\omega}^{\alpha}(\widehat{u}(\widehat{x},\widehat{t}\,))$ satisfy
\begin{equation}
\Psi^{\ast}\widehat{\omega}^{1} = \omega^{1} + d \nu \; ,  \; \; \; \;
\Psi^{\ast}\widehat{\omega}^{2} = \omega^{2}\cosh \nu + \omega^{3} \sinh \nu \; ,  \; \; \; \;
\Psi^{\ast}\widehat{\omega}^{3} = \omega^{2} \sinh \nu + \omega^{3} \cosh \nu \; .
\label{corresp}
\end{equation}
\end{thm}
 
That the maps $\Psi$ and $\nu$ exist, is simply \cite{12,re} an expression of the local 
uniqueness of surfaces of constant curvature referred to above. The proof of Theorem \ref{olya1}
\cite{re} makes it clear that the maps $\Psi$ and $\nu$ depend on {\em both} solutions $u(x,t)$
and $\widehat{u}(\widehat{x},\widehat{t}\,)$. Now, a careful analysis of (\ref{corresp}) allows 
one to find a system of equations for $\gamma$, $\delta$, $\nu$ which can be --in principle-- 
solved without previous knowledge of the solution $\widehat{u}(\widehat{x},\widehat{t}\,)$:
 
\begin{lemma} \label{twogenlem}
Let $\Xi(x,t,u,\dots)=0$ and $\widehat{\Xi}(\widehat{x},\widehat{t},\widehat{u},\dots)=0$ be two
scalar equations describing pseudo-spherical surfaces with associated one--forms $\omega^{i} = 
f_{i 1} dx + f_{i 2}dt$ and $\widehat{\omega}^{i} = \widehat{f}_{i 1}d\widehat{x} + 
\widehat{f}_{i2}d\widehat{t}$, $i=1,2,3$, respectively, in which $\widehat{f}_{11}=\widehat{u}$.
Assume that $\omega^{2} \wedge \omega^{3} \not \equiv 0$, and $\widehat{\omega}^{2} \wedge 
\widehat{\omega}^{3} \not \equiv 0$.

Let $\Upsilon (x,t) = (\gamma(x,t),\delta(x,t))$ be a smooth map with Jacobian 
$J = \gamma_{x}\delta_{t} - \gamma_{t}\delta_{x}$ from (an open subset of) $M$ to (an open subset
of) $\widehat{M}$, and let $\nu$ be a smooth map from (an open subset of) $M$ to ${\bf R}$. The 
system of equations
\begin{eqnarray}
J \, (\Upsilon^{\ast}\widehat{f}_{12}) & = & - \left[ \gamma_{t} (f_{11} + \nu_{x}) - 
\gamma_{x}(f_{12}+ \nu_{t}) \right] \; ,   \\
(\Upsilon^{\ast}\widehat{f}_{21})\,\gamma_{x}+(\Upsilon^{\ast}\widehat{f}_{22})\,\delta_{x}& = &
f_{21}\cosh\nu - f_{31} \sinh\nu \; ,     \\   
(\Upsilon^{\ast}\widehat{f}_{21})\,\gamma_{t}+(\Upsilon^{\ast}\widehat{f}_{22})\,\delta_{t}& = &
f_{22}\cosh\nu - f_{32} \sinh\nu \; ,     \\
(\Upsilon^{\ast}\widehat{f}_{31})\,\gamma_{x}+(\Upsilon^{\ast}\widehat{f}_{32})\,\delta_{x}& = &
f_{21}\sinh\nu + f_{31} \cosh\nu \; ,    \\
(\Upsilon^{\ast}\widehat{f}_{31})\,\gamma_{t}+(\Upsilon^{\ast}\widehat{f}_{32})\,\delta_{t}& = &
f_{22}\sinh\nu + f_{32} \cosh\nu  \; , 
\end{eqnarray}
in which the pull--backs of $\widehat{u}$ and its derivatives with respect to $\widehat{x}, 
\widehat{t}$ appearing in the functions $(\Upsilon^{\ast}\widehat{f}_{ij})(x,t)$ have been 
evaluated by means of the equation
\begin{equation}
\widehat{u}\circ\Upsilon = \frac{1}{J}\left( \delta_{t}(f_{11}+\nu_{x}) - \delta_{x}(f_{12} + 
\nu_{t})\right) \; ,
\end{equation}                  
admits ---whenever $u(x,t)$ is a $I$--generic solution of $\Xi = 0$--- a local solution 
$\gamma(x,t)$, $\delta(x,t)$, $\nu(x,t)$ such that $\Upsilon (x,t) = (\gamma(x,t),\delta(x,t))$ 
is a local diffeomorphism.
\end{lemma}   

It follows that there exists a transformation carrying solutions to $\Xi (x,t,u, \dots ) = 0$ to
solutions to $\widehat{\Xi} (\widehat{x}, \widehat{t}, \widehat{u}, \dots ) = 0$:

\begin{thm}  \label{olya2}
Let $\Xi (x,t,u, \dots ) = 0$ and $\widehat{\Xi} (\widehat{x}, \widehat{t}, \widehat{u}, \dots  
= 0$ be two scalar equations describing pseudo-spherical surfaces with associated one--forms 
$\omega^{\alpha} = f_{\alpha 1}dx + f_{\alpha 2} dt$ and $\widehat{\omega}^{i} = 
\widehat{f}_{\alpha 1}d\widehat{x} + \widehat{f}_{\alpha 2} d\widehat{t}$, $\alpha=1,2,3$, 
respectively, and assume that these one--forms satisfy $\omega^{2} \wedge \omega^{3} \not 
\equiv 0$ and $\widehat{\omega}^{2} \wedge \widehat{\omega}^{3} \not \equiv 0$. 
Any $I$--generic solution $u(x,t)$ of $\Xi (x,t,u, \dots ) = 0$ gives rise to a $I$--generic 
solution $\widehat{u}(\widehat{x},\widehat{t}\,)$ of $\widehat{\Xi} (\widehat{x}, \widehat{t},
\widehat{u}, \dots ) = 0$ by means of
\begin{equation}
\widehat{u}\circ\Psi = \frac{1}{J}\left( \delta_{t}(f_{11}+\nu_{x}) - \delta_{x}(f_{12}+\nu_{t})
    \right),                                                                        \label{a811}
\end{equation}
in which $\nu$ is a real--valued function, $\Psi (x,t) = (\gamma(x,t),\delta(x,t))$ is a local
diffeomorphism, $J$ is the Jacobian of $\Psi$, and both $\nu$ and $\Psi$ are determined by 
$u(x,t)$ by means of Lemma $\ref{twogenlem}$.
\end{thm}   

Note that the local diffeomorphism $\Psi (x,t) = (\gamma(x,t),\delta(x,t))$ and the map $\nu$ 
depend on the solution $u(x,t)$, and therefore Transformation (\ref{a811}) depends on  $u(x,t)$.
This fact shows yet another difference between the correspondences of \cite{12,re} and usual 
B\"{a}cklund transformations. Consider, for example, the classical transformation \cite{BA}
\begin{equation}
\theta_{t} + \omega_{x}  =  - \sin \omega \cos \theta \; , ~ ~ ~ ~ ~ \mbox{ and } ~ ~ ~ ~ ~ 
\theta_{x} + \omega_{t}  =   \cos \omega \sin \theta \;  ,           \label{seno4}
\end{equation}
connecting two solutions of the sine--Gordon equation
\begin{equation}
\frac{\partial^{2}\theta}{\partial x^{2}} - \frac{\partial^{2}\theta}{\partial t^{2}} =
\sin \theta \cos \theta  \; .                      \label{seno}
\end{equation}
In contradistinction with (\ref{a811}), the {\em form} of Equations (\ref{seno4}), which 
determine a ``new'' solution $\omega (x,t)$ to the sine--Gordon equation starting from an 
``old'' solution $\theta (x,t)$, is the same for {\em any} given generic solution $\theta(x,t)$ 
of (\ref{seno}).

One would expect that generalizations of Theorems \ref{olya1} and \ref{olya2} to hierarchies 
of PSS equations should exist: these results are based solely on the local geometry of 
pseudo-spherical surfaces as encoded in the structure equations (\ref{structure0}), and these 
equations also appear when one is dealing with hierarchies!

If $u_{\tau_{i}} = F_{i}$, $i \geq 0$ is a hierarchy of pseudo-spherical type with associated
one--forms $\Theta^{[n]}_{\alpha}$, the pull--backs of the forms $\Theta^{[n]}_{\alpha}$ by 
solutions $u^{[n]}$ to $u_{\tau_{i}} = F_{i}$ will be denoted again by $\Theta^{[n]}_{\alpha}$, 
no confusion should arise. In analogy with the single equation case
considered in \cite{12,re}, one can prove the following result:

\begin{thm} \label{nk}
Let $u_{\tau_{i}} = F_{i}(x,t,u,\dots)$ and $\widehat{u}_{\widehat{\tau}_{i}} = 
\widehat{F}_{i}(\widehat{x},\widehat{t},\widehat{u} ,\dots)$, $i \geq 0$, be two hierarchies of 
pseudo-spherical type with associated one--forms
\begin{equation}
\Theta_{\alpha}^{[n]} = f_{\alpha 1}dx + f_{\alpha 2}dt + \sum_{i=1}^{n} h_{\alpha i} d \tau_{i}
\; , \; \; \; \mbox{  and } \; \; \; 
\widehat{\Theta}_{\alpha}^{[n]} = \widehat{f}_{\alpha 1}d\widehat{x} + 
\widehat{f}_{\alpha 2}d\widehat{t} + \sum_{i=1}^{n} \widehat{h}_{\alpha i} d \widehat{\tau}_{i}
\; ,                                                                                \label{alfa}
\end{equation}
respectively. 
Let $\{u^{[n]}\}$ and $\{\widehat{u}^{[n]}\}$ be solutions of $u_{\tau_{i}} = F_{i}$ and 
$\widehat{u}_{\widehat{\tau}_{i}}=\widehat{F}_{i}$, $i \geq 0$, and assume that
$u^{[0]}(x,t)$ and $\widehat{u}^{[0]}(\widehat{x},\widehat{t}\,)$ are $III$--generic. 
Then, for each $n \geq 0$ there exist a local diffeomorphism $\Upsilon^{[n]} : V^{[n]} 
\rightarrow \widehat{V}^{[n]}$, in which $V^{[n]}$ and $\widehat{V}^{[n]}$ are open subsets of 
the domains of $u^{[n]}$  and $\widehat{u}^{[n]}$ respectively, and a smooth function $\mu^{[n]}
: V^{[n]}\rightarrow {\bf R}$, such that the pull--backs of $\Theta^{[n]}_{\alpha}$ by $u^{[n]}$
and $\widehat{\Theta}^{[n]}_{\alpha}$ by $\widehat{u}^{[n]}$ satisfy
\begin{eqnarray}
\Upsilon^{[n]\ast}\widehat{\Theta}_{1}^{[n]} & = & \Theta_{1}^{[n]}\cos \mu^{[n]} + 
\Theta_{2}^{[n]} \sin \mu^{[n]} \; , \label{II1}\\   
\Upsilon^{[n]\ast}\widehat{\Theta}_{2}^{[n]} & = & -\Theta_{1}^{[n]} \sin \mu^{[n]} + 
\Theta_{2}^{[n]} \cos \mu^{[n]} \; , \label{II2} \\
\Upsilon^{[n]\ast}\widehat{\Theta}_{3}^{[n]} & = & \Theta_{3}^{[n]} + d\mu^{[n]}\; . \label{II3}
\end{eqnarray}
Moreover, the maps $\Upsilon^{[n]}$ and $\mu^{[n]}$, $n \geq 0$, can be chosen so that 
\begin{equation}
\Upsilon^{[n+1]}|_{{V}^{[n]}} = \Upsilon^{[n]} \; , \; \; \; \; \; \; \; \; \mbox{ and } 
\; \; \; \; \; \; \; \; 
\mu^{[n+1]}|_{{V}^{[n]}} = \mu^{[n]} \; , \; \; \; \; \; n \geq 0 \; .       \label{restriction}
\end{equation}
\end{thm}

\begin{proof} 
For each $n \geq 0$, one considers the one--forms $\sigma^{[n]}_{\alpha}$, $\alpha =1,2,3$, 
given by
\begin{equation}
\sigma^{[n]}_{1} = \frac{1}{\widehat{t}} \, d\widehat{x} + \sum_{i=1}^{n} \frac{1}{\widehat{t}} 
\, d \widehat{\tau}_{i} \; , \; \; \; \; \; \; \; \; \; \; 
\sigma^{[n]}_{2} = \frac{1}{\widehat{t}} \, d \widehat{t} \; , 
\; \; \; \; \; \; \; \; \; \; \sigma^{[n]}_{3} = \sigma^{[n]}_{1} \; .
\end{equation}
Note that $\sigma^{[0]}_{1}\otimes\sigma^{[0]}_{1} + \sigma^{[0]}_{2}\otimes\sigma^{[0]}_{2}$ is
exactly the standard hyperbolic metric on the Poincar\'{e} upper half plane, and that 
$\sigma^{[0]}_{3}$ is the corresponding connection one--form. In fact, one easily checks that 
for any $n \geq 0$ the one--forms $\sigma^{[n]}_{\alpha}$ satisfy the structure equations 
\begin{equation}
d \sigma^{[n]}_{1} = \sigma^{[n]}_{3} \wedge \sigma^{[n]}_{2} \; , \; \; \; \; \; \; \; \;
d \sigma^{[n]}_{2} = \sigma^{[n]}_{1} \wedge \sigma^{[n]}_{3} \; , \; \; \; \; \; \; \; \; 
d \sigma^{[n]}_{3} = \sigma^{[n]}_{1} \wedge \sigma^{[n]}_{2} \; , 
\end{equation}
identically. Now, one can ``dress'' these one--forms in the following sense: one can prove that 
for each $n \geq 0$ there exists a local diffeomorphism $\Gamma^{[n]} : (x,t,\tau_{1}, \dots, 
\tau_{n}) \mapsto (\widehat{x},\widehat{t}, \widehat{\tau_{1}},\dots,\widehat{\tau}_{n})$ and a 
real--valued function $\theta^{[n]}(x,t,\tau_{1}, \dots, \tau_{n})$ such that the pull--backs of 
the one--forms $\Theta_{\alpha}^{[n]}$ by the solutions $u^{[n]}$ satisfy 
\begin{eqnarray}
\Gamma^{[n]\ast} \, {\sigma}_{1}^{[n]} & = & \Theta_{1}^{[n]}\cos \theta^{[n]} + \Theta_{2}^{[n]}
\sin \theta^{[n]} \; , \label{pb1}\\   
\Gamma^{[n]\ast} \,{\sigma}_{2}^{[n]} & = & -\Theta_{1}^{[n]} \sin\theta^{[n]} + \Theta_{2}^{[n]}
\cos \theta^{[n]} \; , \label{pb2} \\
\Gamma^{[n]\ast} \, {\sigma}_{3}^{[n]} & = & \Theta_{3}^{[n]} + d \theta^{[n]} \; . \label{pb3}
\end{eqnarray}
This is seen thus: Write $\Gamma^{[n]} = (\alpha,\beta,T_{1}, \dots, T_{n})$, in which
$\alpha, \beta, T_{i} : {V}^{[n]} \rightarrow {\bf R}$, and compute $\Gamma^{[n]\ast} \, 
{\sigma}_{\alpha}^{[n]}$. Equations (\ref{pb1})--(\ref{pb3}) are equivalent to the following 
equations
\begin{eqnarray}
\frac{1}{\beta}(\alpha_{x} + \sum_{i=1}^{n} T_{i,x} ) & = & f_{11} \cos \theta^{[n]} + f_{21} 
\sin \theta^{[n]} \; , \label{pb11} \\
\frac{1}{\beta}(\alpha_{t} + \sum_{i=1}^{n} T_{i,t} ) & = & f_{12} \cos \theta^{[n]} + f_{22} 
\sin \theta^{[n]} \; , \label{pb12} \\
\frac{1}{\beta}(\alpha_{\tau_{j}} + \sum_{i=1}^{n} T_{i,\tau_{j}} ) & = & h_{1j} \cos 
\theta^{[n]} + h_{2j} \sin \theta^{[n]} \; , \; \; \; \; \; \; j =1, \dots n,     \label{pb13}\\
\frac{1}{\beta}\beta_{x} & = & - f_{11} \sin \theta^{[n]} + f_{21} \cos \theta^{[n]} \; , 
                                                                                 \label{pb21} \\
\frac{1}{\beta}\beta_{t} & = & - f_{12} \sin \theta^{[n]} + f_{22} \cos \theta^{[n]},\\   
\frac{1}{\beta}\beta_{\tau_{j}} & = & - h_{1j}\sin \theta^{[n]} + h_{2j} \cos \theta^{[n]} \; , 
\; \; \; \; j =1, \dots n \; , \label{pb23} \\
\frac{1}{\beta}(\alpha_{x} + \sum_{i=1}^{n} T_{i,x} ) & = & f_{31} + \theta^{[n]}_{x} \; , 
                                                                                 \label{pb31} \\
\frac{1}{\beta}(\alpha_{t} + \sum_{i=1}^{n} T_{i,t} ) & = & f_{32} + \theta^{[n]}_{t} \; , 
                                                                                \label{pb32} \\ 
\frac{1}{\beta}(\alpha_{\tau_{j}} + \sum_{i=1}^{n} T_{i,\tau_{j}}) & = & h_{3j} + 
\theta^{[n]}_{\tau_{j}} \; , \; \; \; \;  j =1, \dots n \; .                        \label{pb33}
\end{eqnarray}
Substituting (\ref{pb11})--(\ref{pb13}) into (\ref{pb31})--(\ref{pb33}), one finds that the 
system (\ref{pb11})--(\ref{pb33}) becomes 
\begin{eqnarray}
d (\alpha + \sum_{i=1}^{n} T_{i}) & = & \beta (\Theta_{1}^{[n]} \cos \theta ^{[n]} + 
\Theta_{2}^{[n]} \sin \theta ^{[n]}) \; , \label{pba} \\
d (\ln \beta ) & = & - \Theta_{1}^{[n]} \sin \theta ^{[n]} + \Theta_{2}^{[n]} 
\cos \theta ^{[n]} \; , \label{pbb} \\
\Theta_{1}^{[n]} \cos \theta ^{[n]} + \Theta_{2}^{[n]} \sin \theta ^{[n]} & = & \Theta^{[n]}_{3}
 + d \theta^{[n]} \; . \label{pbc}
\end{eqnarray}
It is easy to see that the Pfaffian system (\ref{pbc}) is completely integrable for a function 
$\theta^{[n]}(x,t,\tau_{1}, \dots, \tau_{n})$, since that for each $n \geq 0$, the structure 
equations
\begin{equation}
d \Theta^{[n]}_{1} = \Theta^{[n]}_{3} \wedge \Theta^{[n]}_{2} \; , \; \; \; \; \; \; \;
d \Theta^{[n]}_{2} = \Theta^{[n]}_{1} \wedge \Theta^{[n]}_{3} \; , \; \; \; \; \; \; \; 
d \Theta^{[n]}_{3} = \Theta^{[n]}_{1} \wedge \Theta^{[n]}_{2} \; ,
\end{equation}
are satisfied on solutions of the hierarchy $u_{\tau_{i}} = F_{i}$, $i \geq 0$. It is also 
straightforward to check that the right hand sides of (\ref{pba}) and
(\ref{pbb}) are closed one--forms. Thus, Equations (\ref{pba}) and (\ref{pbb}) determine $\beta$
and $\alpha + \sum_{i=1}^{n} T_{i}$. One can specify the functions $T_{i}$, $i = 1, \dots, n$ 
almost freely:
they are constrained only by the fact that $\Gamma^{[n]} = (\alpha,\beta,T_{1}, \dots, T_{n})$ be
a local diffeomorphism. A natural choice is to take $T_{i} = \tau_{i}$. It then follows that the
Jacobian determinant of $\Gamma^{[n]}$ is simply 
\[
\alpha_{x} \beta_{t} - \alpha_{t} \beta_{x} = \beta^{2} (f_{11}f_{22} - f_{12} f_{21} ) \; ,
\]
and therefore, since $u^{[0]}(x,t)$ is $III$--generic, $\Gamma^{[n]}$ is a local diffeomorphism.

Next, one can also find a diffeomorphism $\widehat{\Gamma}^{[n]}$ and a function 
$\widehat{\theta}^{[n]}$ satisfying the equations corresponding to Equations 
(\ref{pb1})--(\ref{pb3}) with $\Theta_{\alpha}^{[n]}$
replaced by $\widehat{\Theta}_{\alpha}^{[n]}$. It then follows that 
\[
\Upsilon^{[n]} = (\widehat{\Gamma}^{[n]})^{-1} \circ \Gamma^{[n]}, \; \; \; \; \mbox{ and } 
\; \; \; \; 
\mu^{[n]} = \theta^{[n]} - \widehat{\theta}^{[n]} \circ \Upsilon^{[n]}
\]
satisfy Equations (\ref{II1})--(\ref{II3}). 

It remains to check compatibility of the functions $\Upsilon^{[n]}$ and $\mu^{[n]}$, $n \geq 0$.
The fact that $\Upsilon^{[n+1]}|_{{V}^{[n]}} = \Upsilon^{[n]}$ follows trivially from the 
construction of the functions $\Upsilon^{[\cdot]}$. On the other hand, in order to check that 
$\mu^{[n+1]}|_{{V}^{[n]}} = \mu^{[n]}$, it is of course enough to show that 
\begin{equation}
\theta^{[n+1]}|_{{V}^{[n]}} = \theta^{[n]} \; , \; \; \; \; \; n \geq 0 \; .  \label{rest0}
\end{equation} 
Expanding (\ref{pbc}) (with $\theta^{[n]}$ replaced by $\theta^{[n+1]}$) one sees that the 
function $\theta^{[n+1]}$ is determined by the Pfaffian system
\begin{eqnarray}
\Theta_{1}^{[n]} \cos \theta ^{[n+1]} + \Theta_{2}^{[n]} \sin \theta ^{[n+1]} & = & 
\Theta^{[n]}_{3} + d \, \theta^{[n+1]} \; ,                                      \label{rest1} \\
h_{1 , n+1} \cos \theta^{[n+1]} + h_{2 , n+1} \sin \theta^{[n+1]} & = &  h_{3 , n+1} + 
\frac{\partial \, \theta^{[n+1]}}{\partial \tau_{n+1}} \; ,   \label{rest2}
\end{eqnarray}
in which the operator $d$ appearing in (\ref{rest1}) is the exterior derivative on the space 
${\bf R}^{n+2}$ equipped with coordinates $(x,t,\tau_{1}, \dots , \tau_{n})$. One now considers 
the following Cauchy problem:
\[
\left\{
\begin{array}{rcl}
\displaystyle 
\frac{\partial \, z}{\partial \tau_{n+1}} & = & h_{1,n+1} \cos z + h_{2,n+1} \sin z - h_{3,n+1} 
                                                                                         \; ,  \\
z(x,t,\tau_{1}, \dots , \tau_{n}, 0) & = & \theta^{[n]}(x,t,\tau_{1}, \dots , \tau_{n}) \; .
\end{array}
\right.
\]
This problem has a unique solution $z(x,t,\tau_{1}, \dots , \tau_{n}, \tau_{n+1})$, see Gardner 
\cite{Ga}. It is then enough to set $\theta^{[n+1]} = z$. Since $\theta^{[n]}$ satisfies Equation
(\ref{rest1}) by construction, the function $\theta^{[n+1]}$ satisfies (\ref{rest1}), 
(\ref{rest2}), and $\theta^{[n+1]}|_{{V}^{[n]}} = \theta^{[n]}$. This ends the proof. 
\end{proof}

Theorem \ref{nk} is already a correspondence result, as the following corollary shows:

\begin{cor} \label{cor1}
Let $u_{\tau_{i}} = F_{i}(x,t,u,\dots)$ and $\widehat{u}_{\widehat{\tau}_{i}}=\widehat{F}_{i}
(\widehat{x}, \widehat{t},\widehat{u} ,\dots)$, $i \geq 0$, be two hierarchies of 
pseudo-spherical type with associated one--forms given by $(\ref{alfa})$. Let $\{u^{[n]}\}$ and 
$\{\widehat{u}^{[n]}\}$ be solutions of $u_{\tau_{i}} = F_{i}$ and 
$\widehat{u}_{\widehat{\tau}_{i}}=\widehat{F}_{i}$, $i \geq 0$, and assume 
that $u^{[0]}(x,t)$ and $\widehat{u}^{[0]}(\widehat{x},\widehat{t}\,)$ are $III$--generic. There
exist sequences of maps $\{ \Upsilon^{[n]} \}$ and $\{ \mu^{[n]} \}$ such that: 

$(a)$ for each $n \geq 0$, $\Upsilon^{[n]} = (\psi^{[n]},\varphi^{[n]}, \phi_{1}^{[n]},\dots,
\phi_{n}^{[n]})$ is a local diffeomorphism from an open subset $V^{[n]}$ of the domain of 
$u^{[n]}$ to an open subset $\widehat{V}^{[n]}$ of the domain of $\widehat{u}^{[n]}$;

$(b)$ for each $n \geq 0$, $\mu^{[n]} : V^{[n]} \rightarrow {\bf R}$ is a smooth real--valued 
function; and 

$(c)$ for each $n \geq 0$, the pull--backs of the functions $f_{\alpha\beta}$, $h_{\alpha i}$,
$\widehat{f}_{\alpha\beta}$, and $\widehat{h}_{\alpha i}$ by $u^{[n]}$ and $\widehat{u}^{[n]}$ 
respectively, satisfy
\begin{equation}
\widehat{f}_{11} \circ \Upsilon^{[n]} = \frac{\Delta_{1}^{[n]}}{\Delta^{[n]}} \; , 
									~ ~ ~ ~ ~ ~ ~ ~ ~ ~ 
\widehat{f}_{12} \circ \Upsilon^{[n]} = \frac{\Delta_{2}^{[n]}}{\Delta^{[n]}} \; , 
									~ ~ ~ ~ ~ ~ ~ ~ ~ ~ 
\widehat{h}_{1j} \circ \Upsilon^{[n]} = \frac{\Delta_{j+2}^{[n]}}{\Delta^{[n]}} \; ,
\end{equation}
in which
\begin{equation}
\Delta^{[n]} = \left| 
\begin{array}{ccccc}
\psi_{x}^{[n]} & \varphi_{x}^{[n]} & 0 & \dots & 0 \\
\psi_{t}^{[n]} & \varphi_{t}^{[n]} & 0 & \dots & 0 \\
\psi_{\tau_{1}}^{[n]} & \varphi_{\tau_{1}}^{[n]} & 1 & \dots & 0 \\
\vdots & \vdots & \vdots & \vdots & \vdots \\
\psi_{\tau_{n}}^{[n]} & \varphi_{\tau_{n}}^{[n]} & 0 & \dots & 1 
\end{array}
\right| 
= \psi_{x}^{[n]} \varphi_{t}^{[n]} - \varphi_{x}^{[n]} \psi_{t}^{[n]} \; ,      \label{delta}
\end{equation} 
and for each $i =1,2,\dots n+2$, $\Delta_{i} = \Delta$, except for the $i$th column, which is 
replaced by the column vector 
\begin{equation}
\left( 
\begin{array}{c}
f_{11}\cos \mu^{[n]} + f_{21} \sin \mu^{[n]} \\
f_{12}\cos \mu^{[n]} + f_{22} \sin \mu^{[n]} \\
h_{11}\cos \mu^{[n]} + h_{21} \sin \mu^{[n]} \\
\vdots                                       \\
h_{1n}\cos \mu^{[n]} + h_{2n} \sin \mu^{[n]} 
\end{array}
\right) \; .                                                 \label{delta1}
\end{equation}
\end{cor}
\begin{proof} 
Theorem \ref{nk} implies that there exist sequences $\{ \Upsilon^{[n]} \}$ and $\{ \mu^{[n]} \}$ 
of local diffeomorphisms and smooth functions respectively, such that for each $n \geq 0$, 
\begin{eqnarray}
\Upsilon^{[n]\ast}\widehat{\Theta}_{1}^{[n]} & = & \Theta_{1}^{[n]}\cos \mu^{[n]} + 
\Theta_{2}^{[n]} \sin \mu^{[n]} \; ,                                            \label{II11}\\   
\Upsilon^{[n]\ast}\widehat{\Theta}_{2}^{[n]} & = & -\Theta_{1}^{[n]} \sin \mu^{[n]} + 
\Theta_{2}^{[n]} \cos \mu^{[n]} \; ,                                              \label{II21} \\
\Upsilon^{[n]\ast}\widehat{\Theta}_{3}^{[n]} & = & \Theta_{3}^{[n]} + d \mu^{[n]} \; . 
                                                                                   \label{II31}
\end{eqnarray}
Write $\Upsilon^{[n]} = (\psi^{[n]},\varphi^{[n]},\phi_{1}^{[n]},\dots,\phi_{n}^{[n]})$, and set
\begin{equation}
\widehat{g}_{\alpha\beta} = \widehat{f}_{\alpha\beta}  \circ \Upsilon^{[n]}\; , \; \; \; \; \; \;
\widehat{H}_{\alpha j} =  \widehat{h}_{\alpha j} \circ \Upsilon^{[n]} \; .           \label{aux}
\end{equation}
The construction of $\Upsilon^{[n]}$ appearing in the proof of Theorem 4 implies that
\[
\phi_{j}^{[n]} = \tau_{j}, \; \; \; \; \; \; \; \; \; \;  j = 1,2,\dots, n \; .
\]
Thus, the system (\ref{II11})--(\ref{II31}) becomes
\begin{eqnarray}
\widehat{g}_{11}\;\psi_{x}^{[n]} + \widehat{g}_{12}
\;\varphi_{x}^{[n]} & = & f_{11} \cos \mu^{[n]} + f_{21} \sin \mu^{[n]}              \label{a} \\
\widehat{g}_{11}\;\psi_{t}^{[n]} + \widehat{g}_{12}
\;\varphi_{t}^{[n]} & = & f_{12} \cos \mu^{[n]} + f_{22} \sin \mu^{[n]}              \label{b} \\
\widehat{g}_{11}\;\psi_{\tau_{j}}^{[n]} + \widehat{g}_{12}
\;\varphi_{\tau_{j}}^{[n]} + \widehat{H}_{1j} & = & h_{1j} 
\cos \mu^{[n]} + h_{2j} \sin \mu^{[n]}                                             \label{cc} \\
\widehat{g}_{21}\;\psi_{x}^{[n]} + \widehat{g}_{22}
\;\varphi_{x}^{[n]} & = & -f_{11} \sin \mu^{[n]} + f_{21} \cos \mu^{[n]}             \label{d} \\
\widehat{g}_{21}\;\psi_{t}^{[n]} + \widehat{g}_{22}
\;\varphi_{t}^{[n]} & = & - f_{12} \sin \mu^{[n]} + f_{22} \cos \mu^{[n]}            \label{e} \\
\widehat{g}_{21}\;\psi_{\tau_{j}}^{[n]} + \widehat{g}_{22}
\;\varphi_{\tau_{j}}^{[n]} + \widehat{H}_{2j} & = & -h_{2j} 
\sin \mu^{[n]} + h_{2j} \cos \mu^{[n]}                                               \label{f} \\
\widehat{g}_{31}\;\psi_{x}^{[n]} + \widehat{g}_{32}
\;\varphi_{x}^{[n]} & = & f_{31} + \mu_{x}^{[n]}                                     \label{g} \\
\widehat{g}_{31}\;\psi_{t}^{[n]} + \widehat{g}_{32}
\;\varphi_{t}^{[n]} & = & f_{32} + \mu_{t}^{[n]}                                     \label{h} \\
\widehat{g}_{31}\;\psi_{\tau_{j}}^{[n]} + \widehat{g}_{32} 
\;\varphi_{\tau_{j}}^{[n]} + \widehat{H}_{3j} & = & h_{3j} + 
\mu_{\tau_{j}}^{[n]}                                                                   \label{c}
\end{eqnarray}
in which $j = 1,2, \dots , n$. Consider the $n+2$ equations (\ref{a})--(\ref{cc}) as a linear 
system for $\widehat{g}_{11}$, $\widehat{g}_{12}$, and $\widehat{H}_{1j}$, $j = 1,2, \dots , n$.
It is easy to see that the determinant of (\ref{a})--(\ref{cc}) is given by (\ref{delta}), and 
therefore Cramer's rule implies that 
\[
\widehat{g}_{11} = \widehat{f}_{11}\circ\Upsilon^{[n]}=\frac{\Delta_{1}^{[n]}}{\Delta^{[n]}}\; , 
							~ ~ ~ ~ ~ ~ ~ ~ ~ ~ 
\widehat{g}_{12} = \widehat{f}_{12}\circ\Upsilon^{[n]}=\frac{\Delta_{2}^{[n]}}{\Delta^{[n]}}\; , 
									~ ~ ~ ~ ~ ~ ~ ~ ~ ~ 
\widehat{H}_{1j} = \widehat{h}_{1j}\circ\Upsilon^{[n]}=\frac{\Delta_{j+2}^{[n]}}{\Delta^{[n]}}\;,
\]
in which $\Delta_{i}$ is determined by (\ref{delta}) and (\ref{delta1}).
\end{proof}

Next, one would like to use Theorem \ref{nk} to {\em construct} a B\"{a}cklund--like 
transformation from $u^{[n]}$ to $\widehat{u}^{[n]}$, $n \geq 0$. In order to avoid some 
technicalities related to the gauge freedom one has to determine one--forms associated
with differential equations (see Remark \ref{gauge} and \cite{re}) only hierarchies of strictly 
pseudo-spherical  type will be considered in what follows. Motivated by Corollary 1, let
\begin{equation}
\Upsilon^{[n]} = (\psi^{[n]},\varphi^{[n]},\phi_{1}^{[n]},\dots, \phi_{n}^{[n]}) \; ,  
\; \; \; \; \; \; \; n \geq 0 \; ,  \label{map} 
\end{equation}
be a map from an open subset of ${\bf R}^{n+2}$ with coordinates $(x ,t,\tau_{1},\dots,\tau_{n})$
to an open subset of ${\bf R}^{n+2}$ with coordinates 
$(\widehat{x},\widehat{t},\widehat{\tau}_{1},\dots, \widehat{\tau}_{n})$, and let 
$\mu (x ,t,\tau_{1},\dots,\tau_{n})$ be a smooth real--valued function on an open subset of 
${\bf R}^{n+2}$. Also, define determinants $\Delta^{[n]}$ and $\Delta^{[n]}_{i}$ as
\begin{equation}
\Delta^{[n]} = \mbox{Jacobian}(\Upsilon^{[n]}) \; ,
\end{equation}
and $\Delta^{[n]}_{i} = \Delta^{[n]}$, except for the $i$th column, which is replaced by the 
vector (\ref{delta1}). As in the single equation case \cite{12,re} one now proves the 
following technical lemma:
 
\begin{lemma} \label{lena0}
Let $u_{\tau_{i}} = F_{i}$ and $\widehat{u}_{\widehat{\tau}_{i}} = \widehat{F}_{i}$ be two 
hierarchies of strictly pseudo-spherical type with associated functions $f_{\alpha\beta}$, 
$h_{\alpha i}$, and $\widehat{f}_{\alpha\beta}$, $\widehat{h}_{\alpha i}$ respectively, and 
assume that $\widehat{f}_{11} = G(\widehat{u})$, $G' \neq 0$. For each $n \geq 0$, let 
$\Upsilon^{[n]}$ be a smooth map as in $(\ref{map})$, and let $\mu^{[n]}(x ,t,\tau_{1},\dots,
\tau_{n})$ be a smooth real--valued function on an open subset of ${\bf R}^{n+2}$. Set 
$\widehat{g}_{\alpha\beta} = \widehat{f}_{\alpha\beta} \circ\Upsilon^{[n]}$ and 
$\widehat{H}_{\alpha j} = \widehat{h}_{\alpha j} \circ \Upsilon^{[n]}$. The system of equations
\begin{eqnarray}
\Delta^{[n]} \; \widehat{g}_{12} & = & \Delta_{2}^{[n]}                        \label{a11} \\
\Delta^{[n]} \; \widehat{H}_{1j} & = & \Delta_{j+2}^{[n]}            \\
\widehat{g}_{21}\;\psi_{x}^{[n]} + \widehat{g}_{22}\;
\varphi_{x}^{[n]} + \sum_{i=1}^{n} \widehat{H}_{2i}\;\phi_{i,x} 
& = & f_{21} \cos \mu^{[n]} - f_{11} \sin \mu^{[n]}  \; \; \; \; \; \; \; \, \\
\widehat{g}_{21}\;\psi_{t}^{[n]} + \widehat{g}_{22}\;
\varphi_{t}^{[n]} + \sum_{i=1}^{n} \widehat{H}_{2i}\; \phi_{i,t} 
& = & f_{22} \cos \mu^{[n]} - f_{12} \sin \mu^{[n]}   \; \; \; \; \; \; \; \, \\
\widehat{g}_{21}\;\psi_{\tau_{j}}^{[n]} + \widehat{g}_{22}\;
\varphi_{\tau_{j}}^{[n]} + \sum_{i=1}^{n} \widehat{H}_{2j}\; \phi_{i,\tau_{j}}
 & = & h_{2j} \cos \mu^{[n]} - h_{2j} \sin \mu^{[n]}   \; \; \; \; \; \; \;   \\
\widehat{g}_{31}\;\psi_{x}^{[n]} + \widehat{g}_{32}
\varphi_{x}^{[n]} + \sum_{i=1}^{n} \widehat{H}_{3i} \; \phi_{i,x}
& = & f_{31} + \mu_{x}^{[n]}  \\
\widehat{g}_{31}\;\psi_{t}^{[n]} + \widehat{g}_{32}\;
\varphi_{t}^{[n]} + \sum_{i=1}^{n} \widehat{H}_{3i}\; \phi_{i,t} 
& = & f_{32} + \mu_{t}^{[n]}  \\
\widehat{g}_{31}\;\psi_{\tau_{j}}^{[n]} + \widehat{g}_{32}\;
\varphi_{\tau_{j}}^{[n]} + \sum_{i=1}^{n} \widehat{H}_{3j}\; \phi_{i,\tau_{j}} 
& = & h_{3j} + \mu_{\tau_{j}}^{[n]} \; ,                                          \label{a51} 
\end{eqnarray}
in which $j = 1,2, \dots n$, and the left hand side of 
$(\ref{a11})$--$(\ref{a51})$ has been evaluated by means of the equation
\begin{equation}
\widehat{f}_{11} \circ \Upsilon^{[n]} = G(\widehat{u}) \circ \Upsilon^{[n]} = 
\frac{\Delta_{1}^{[n]}}{\Delta^{[n]}} \; ,   \label{a8}
\end{equation}
admits ---whenever $\{u^{[n]}\}$ is a solution of the hierarchy $u_{\tau_{i}} = F_{i}$ such that 
$u^{[0]}(x,t)$ is III--generic--- a local solution $\psi^{[n]}$, $\varphi^{[n]}$, 
$\phi_{1}^{[n]}$, ... $\phi_{n}^{[n]}$, and $\mu^{[n]}$ such that the map $\Upsilon^{[n]} = 
(\psi^{[n]}, \varphi^{[n]}, \phi_{1}^{[n]}, \dots , \phi_{n}^{[n]})$ is a local diffeomorphism. 
Moreover, the maps $\Upsilon^{[n]}$ and $\mu^{[n]}$, $n \geq 0$, can be chosen so that
\[
\Upsilon^{[n+1]}|_{{V}^{[n]}} = \Upsilon^{[n]}, \; \; \; \; \; \; \; \; \mbox{ and } 
\; \; \; \; \; \; \; \; 
\mu^{[n+1]}|_{{V}^{[n]}} = \mu^{[n]}, \; \; \; \; \; n \geq 0, 
\]
in which ${V}^{[n]}$ is the domain of $\Upsilon^{[n]}$ and $\mu^{[n]}$.
\end{lemma}
\begin{proof}
Take {\em any} solution of the hierarchy $\widehat{u}_{\widehat{\tau}_{i}} = 
\widehat{F}_{i}$, $\{ \widehat{u}^{[n]} \}$ say, such that $\widehat{u}^{[0]}(x,t)$ is 
III--generic. By Theorem 4, there exist functions $\psi^{[n]}$, $\varphi^{[n]}$, $\psi_{i}^{[n]}$
and  $\mu^{[n]}$ satisfying the first order system of equations (\ref{II1})--(\ref{II3}), and 
such that
\[
\Upsilon^{[n]} = (\psi^{[n]}, \varphi^{[n]}, \phi_{1}^{[n]}, \dots , \phi_{n}^{[n]})
\]
is a local diffeomorphism. These functions also satisfy the system of
equations (\ref{a11})--(\ref{a51}). Indeed, Corollary \ref{cor1} yields Equation (\ref{a8}) and 
the equations 
\begin{equation}
\widehat{f}_{12} \circ \Upsilon^{[n]} = \Delta^{[n]}_{2}/\Delta^{[n]}, ~ ~ \; \; \;  
										\mbox{ and } ~ ~ 
\; \; \; \widehat{h}_{1j} \circ \Upsilon^{[n]} = \Delta^{[n]}_{j+2}/\Delta^{[n]} .   \label{pul}
\end{equation}
Now, let $\widehat{g}_{\alpha\beta}$, and $\widehat{H}_{\alpha j}$, $\alpha =1,2,3$, be the 
functions depending on $x$, $t$, $\psi^{[n]}$, $\varphi^{[n]}$, $\phi_{1}^{[n]}$, ..., 
$\phi_{n}^{[n]}$, and their derivatives, which are obtained from $\widehat{f}_{\alpha\beta}$ and
$\widehat{h}_{\alpha j}$ by computing the pull--backs $\Upsilon^{\ast}\widehat{f}_{\alpha\beta}$
and $\Upsilon^{\ast}\widehat{h}_{\alpha j}$ as in the enunciate of the lemma. Then, on solutions
of the system (\ref{II11})--(\ref{II31}), one has, for any $\alpha$ and $\beta$,
\[
\left( \widehat{g}_{\alpha\beta} \right) (x,t) = 
                                 \widehat{f}_{\alpha\beta}  \circ \Upsilon (x,t)^{[n]},
\]
since on these solutions, Equation (\ref{a8}) is an identity.
 
Thus, the system (\ref{a11})--(\ref{a51}) reduces to the first order system 
(\ref{II11})--(\ref{II31}), (\ref{pul}), and the result follows. 
\end{proof}
 
Theorem \ref{nk} and Lemma \ref{lena0} allow one to prove the following correspondence result:
 
\begin{thm}  \label{lena1}
Let $u_{\tau_{i}} = F_{i}$ and $\widehat{u}_{\widehat{\tau}_{i}} = \widehat{F}_{i}$ be two 
hierarchies of strictly pseudo-spherical type with associated functions $f_{\alpha\beta}$, 
$h_{\alpha i}$, and $\widehat{f}_{\alpha\beta}$, $\widehat{h}_{\alpha i}$ respectively. Assume 
that $\widehat{f}_{11} = G(\widehat{u})$, $G' \neq 0$. Let $\Upsilon^{[n]}$ be a solution as in 
$(\ref{map})$ of the system of equations $(\ref{a11})$--$(\ref{a51})$ which is a local 
diffeomorphism. Then, any solution $\{u^{[n]}\}$ of the hierarchy $u_{\tau_{i}} = F_{i}$ such 
that $u^{[0]}(x,t)$ is III--generic, determines a solution $\{\widehat{u}^{[n]}\}$ of the 
hierarchy $\widehat{u}_{\widehat{\tau}_{i}} = \widehat{F}_{i}$ such that 
$\widehat{u}^{[0]}(x,t)$ is also III--generic.
\end{thm}
\begin{proof}
Fix a solution $\{u^{[n]} \}$ of the hierarchy $u_{\tau_{i}} = F_{i}$ such that $u^{[0]}(x,t)$ is
III--generic, and consider the system of equations (\ref{a11})--(\ref{a51}). By Lemma 3, this 
system possesses local solutions $\varphi^{[n]}$, $\psi^{[n]}$ and $\mu^{[n]}$ such that 
$\Upsilon^{[n]} = (\psi^{[n]},\varphi^{[n]} , \phi_{1} \dots, \phi_{n})$ is a local 
diffeomorphism with domain $V^{[n]}$, say. One then defines $\widehat{u}^{[n]} \circ 
\Upsilon^{[n]}$ by means of 
\begin{equation}
\widehat{f}_{11} \circ \Upsilon^{[n]} = \frac{\Delta_{1}^{[n]}}{\Delta^{[n]}} \; .   \label{a10}
\end{equation}
Note that 
\[
\widehat{u}^{[n+1]} \circ \Upsilon^{[n+1]}|_{V^{[n]}} = \widehat{u}^{[n]}\circ \Upsilon^{[n]}\; ,
\]
and that for each $n \geq 0$ one obtains a system of six equations equivalent to
\begin{eqnarray}
\Upsilon^{[n]\ast}\widehat{\Theta}_{1}^{[n]} & = & \Theta_{1}^{[n]} \cos \mu^{[n]} + 
\Theta_{3}^{[n]} \sin \mu^{[n]}, \label{II111}\\
\Upsilon^{[n]\ast}\widehat{\Theta}_{2}^{[n]} & = & -\Theta_{1}^{[n]} \sin \mu^{[n]} + 
\Theta_{2}^{[n]} \cos \mu^{[n]} , \label{II211}  \\
\Upsilon^{[n]\ast}\widehat{\Theta}_{3}^{[n]} & = & \Theta_{3}^{[n]} + d \mu^{[n]} . \label{II311}
\end{eqnarray}
 
Since $\Upsilon^{[n]}$ is a local diffeomorphism, and the one--forms $\Theta_{i}^{[n]}$ satisfy 
the structure equations (\ref{109}), so do the one--forms $\widehat{\Theta}_{i}^{[n]}$. 
This means that (\ref{a10}) determines a solution of the hierarchy 
$\widehat{u}_{\widehat{\tau}_{i}} = \widehat{F}_{i}$, as claimed. Finally, note that   
\begin{equation}
\left[ (\widehat{f}_{11}\circ\Upsilon)(\widehat{f}_{32}\circ\Upsilon) -
(\widehat{f}_{12}\circ\Upsilon)(\widehat{f}_{31}\circ\Upsilon) \right] \left[\phi_{x}\psi_{t} -
\phi_{t}\psi_{x} \right] = f_{11}f_{32} - f_{12}f_{31} \; , \label{a9}
\end{equation}
so that $\widehat{u}^{[0]}(\widehat{x},\widehat{t}\,)$ really is a $III$--generic solution. 
\end{proof}

Thus, solutions to hierarchies of pseudo-spherical type --and correspondences between two 
such-- have been found in this paper basically by ``dressing'' a standard hierarchy of 
pseudo-spherical structures ---the one naturally induced by the Poincar\'{e} metric--- by means 
of Equation (\ref{I}) of Proposition 2. One obviously expects that results analogous to the ones
proven in this section may be obtained by considering now transformations (\ref{II}) and 
(\ref{III}) of Proposition 2, as in \cite{re}. 

It is important to note that, as in the single equation case \cite{12,re}, the 
transformation (\ref{a10}) appearing in Theorem \ref{lena1} depends on the particular solution 
$\{u^{[n]}\}$ of the hierarchy $u_{\tau_{i}} = F_{i}$ one starts with, and that its construction
relies strongly on the Frobenius theorem for Pfaffian systems. Since there is no Frobenius 
theorem available at the level of equation manifolds \cite{16} Theorem \ref{lena1} does not 
imply that one can ``transfer'' information on, for instance, conservation laws, from one 
hierarchy to another by means of (\ref{a10}). 

\begin{remark}
Instances of correspondences between solutions of differential equations have 
been also found in the framework of hierarchies of zero curvature equations constructed from 
affine (twisted or untwisted) Ka\v{c}--Moody algebras, by Ferreira, Miramontes, and S\'{a}nchez 
Guill\'{e}n \cite{FMS}. It would be interesting to study whether their approach can be related 
to the one presented here.
\end{remark}

The paper ends with an illustration of Theorems \ref{nk} and \ref{lena1}:

\paragraph{Example: Straightening--out the KdV hierarchy.}

It is shown in this example how one can pass from a solution of the KdV hierarchy to a solution 
of a hierarchy of linear equations. To start with, it follows from the seminal paper \cite{CP} 
by S.S. Chern and C.K. Peng, that the KdV hierarchy $u_{\tau_{i}} = F_{i}$, $i \geq 0$, is of 
strictly pseudo-spherical type with associated functions  
\begin{eqnarray} \label{functions-kdv}
f_{11} & = & 1 - u \; ; \\
f_{21} & = & \lambda \; ; \\ 
f_{31} & = & - 1 - u \; ; \\
h_{1i} & = & (1/2) \lambda B_{x}^{(i+1)} - (1/2) B_{xx}^{(i+1)} - u\,B^{(i+1)} + B^{(i+1)}\; ; \\
h_{2i} & = & \lambda B^{(i+1)} - B^{(i+1)}_{x} \; ; \\
h_{3i} & = & (1/2) \lambda B_{x}^{(i+1)} - (1/2) B_{xx}^{(i+1)} - u\,B^{(i+1)} - B^{(i+1)} \; ,
\end{eqnarray}
in which 
\[
B^{(i)} = \sum_{j=0}^{i} B_{j} \, \lambda^{2(i-j)} \; ,
\]
and the differential functions $B_{j}$ are defined recursively by means of
\begin{eqnarray}
B_{0,x} & = & 0 \, , \\
B_{j+1,x} & = & B_{j,xxx} + 4 u B_{j,x} + 2 u_{x}B_{j} \; , \; \; \; \; 0 \leq j \leq i -1 \; .
\end{eqnarray}
The functions $F_{i}$, $i \geq 0$ are given by
\[
F_{i} = (1/2) B_{i+1,xxx} + u_{x}B_{i+1} + 2 u B_{i+1,x} \; ,
\]
and for instance, one can easily check that the equation $u_{\tau_{0}} = F_{0}$ the standard KdV
equation $u_{t} = u_{xxx} + 6 u u_{x}$.

On the other hand, consider the hierarchy $\widehat{u}_{\widehat{\tau}_{i}} = \widehat{F}_{i}$, 
$i \geq 0$, with seed equation $\widehat{u}_{\widehat{t}} = \widehat{u}_{\widehat{x}\widehat{x}}
 + \widehat{u}_{\widehat{x}}$, and higher equations
\begin{equation}
\widehat{u}_{\widehat{\tau}_{i}}  =
a^{i+2}_{i+1}\,\widehat{u}_{\widehat{x}^{i+2}} + \sum_{l=1}^{i}a^{i+2}_{l}\,
\widehat{u}_{\widehat{x}^{l+1}} + \sum_{l=1}^{i+1}a^{i+2}_{l}\,\widehat{u}_{\widehat{x}^{l}} \; ,
								    \label{lineareqn}
\end{equation}
in which the constants $a^{r}_{s}$ are arbitrary except that $a^{r}_{1} = 1$, $r \geq 1$.
It is straightforward to check that the hierarchy (\ref{lineareqn}) is of strictly 
pseudo-spherical type with associated functions 
\begin{eqnarray} \label{functions-linear}
\widehat{f}_{11} & = & \widehat{u} \; ; \\
\widehat{f}_{21} & = & 1 \; ; \\
\widehat{f}_{31} & = & \widehat{u} \; ; \\
\widehat{h}_{1i} & = & \sum_{k=1}^{i+1} a^{i+2}_{k}\,\widehat{u}_{\widehat{x}^{k}} \; ; \\
\widehat{h}_{2i} & = & 0 \; ; \\
\widehat{h}_{1i} & = & \widehat{h}_{1i} \; .
\end{eqnarray}

Now, the proof of Theorem \ref{nk} implies that if $\{u^{[n]}\}$, $n \geq 0$, is a
solution of the KdV hierarchy such that $u^{[0]}(x,t)$ is $III$--generic, 
there exist functions $\alpha (x,t,\dots , \tau_{n})$, $\beta(x,t,\dots , \tau_{n})$ 
and $\theta^{[n]}(x,t,\dots , \tau_{n})$, $n \geq 0$, such that Equations 
(\ref{pba})--(\ref{pbc}) hold. One then defines a diffeomorphism 
$\Upsilon : (x,t,\tau_{1},\dots,\tau_{n}) \mapsto (\widehat{x},\widehat{t},
\widehat{\tau}_{1},\dots,\widehat{\tau}_{n})$ by means of
\begin{eqnarray}
\widehat{x} & = & - \ln \left| \beta + \frac{1}{\beta}(\alpha + \sum_{i=1}^{n}\tau_{i})^{2} 
                                                                                       \right| \\
\widehat{t} & = & - \left( \frac{\alpha + \sum_{i=1}^{n}\tau_{i}}{\beta} \right) +
1 - \sum_{i=1}^{n}\tau_{i} + \ln \left| \beta + \frac{1}{\beta}(\alpha + 
\sum_{i=1}^{n}\tau_{i})^{2} \right| \\ 
\widehat{\tau}_{i} & = & \tau_{i} \; , 
\end{eqnarray}
in which $i \geq 1$. Theorem \ref{lena1} implies that the sequence 
$\{ \widehat{u}^{[n]} \}_{n \geq 0}$ in which
\[
\widehat{u}^{[n]} = \widehat{x} + \widehat{t} + \widehat{\tau}_{1} +
\dots + \widehat{\tau_{n}} \; , 
\]
is a solution to the hierarchy (\ref{lineareqn}).

\paragraph{\bf Acknowledgements:} Most of this paper was written while the author was at Yale 
University as a Postdoctoral Fellow of the Natural Sciences and Engineering Research Council 
of Canada. He is most grateful to Prof. R. Beals for his help in making the author's stay at 
Yale possible, and for enlightening conversations on the topic of this work. He also thanks 
Profs. L.A. Dickey, P.J. Olver and G. Walschap for making several important suggestions.

\end{document}